\PassOptionsToPackage{table}{xcolor}
\documentclass[fleqn,10pt]{wlscirep}
\usepackage[utf8]{inputenc}
\usepackage[T1]{fontenc}
\usepackage{multirow}
\usepackage{tcolorbox}
\usepackage[table]{xcolor}
\tcbuselibrary{breakable}
\usepackage{booktabs}
\usepackage[table]{xcolor}
\usepackage{graphicx,epstopdf}
\usepackage{amsmath,amsthm,amssymb,amsfonts}
\usepackage{tabularx, pdfcomment}
\usepackage{graphicx, color, multirow,amsmath}
\usepackage{cite}
\usepackage{makecell}
\usepackage{hyperref}
\usepackage[square,sort,comma,numbers]{natbib}
\usepackage{comment}
\usepackage{subfigure}
\usepackage{hyperref}
\usepackage{url}
\usepackage{cases}
\usepackage{bbm}
\usepackage{bm}
\usepackage{url}
\usepackage{algpseudocode}
\usepackage{algorithmicx}
\usepackage{algorithm}
\usepackage{graphicx} 
\usepackage{authblk}
\usepackage[utf8]{inputenc}
\usepackage[T1]{fontenc}
\usepackage{multirow}
\usepackage{tcolorbox}
\tcbuselibrary{breakable}

\usepackage{longtable}
\usepackage{booktabs}

\usepackage{hyperref}

\title{Crystalline Material Discovery in the Era of Artificial Intelligence}

\author[1,*]{Zhenzhong Wang}
\author[1,*]{Haowei Hua}
\author[1,2,$\dag$]{Wanyu Lin}
\author[3]{Ming Yang}
\author[2]{Kay Chen Tan}
\affil[1]{The Hong Kong Polytechnic University, Department of Computing}
\affil[2]{The Hong Kong Polytechnic University, Department of Data Science and Artificial Intelligence}
\affil[3]{The Hong Kong Polytechnic University, Department of Applied Physics}

\affil[$\dag$]{Corresponding author. Email: wan-yu.lin@polyu.edu.hk;}
\affil[*]{These authors contributed equally to this work.}

\begin{abstract}
 Crystalline materials, with symmetrical and periodic structures, exhibit a wide spectrum of properties and have been widely used in numerous applications across electronics, energy, and beyond. For crystalline materials discovery, traditional experimental and computational approaches are time-consuming and expensive. In these years, thanks to the explosive amount of crystalline materials data, great interest has been given to data-driven materials discovery. Particularly, recent advancements have exploited the expressive representation ability of deep learning to model the highly complex atomic systems within crystalline materials, opening up new avenues for efficient and accurate materials discovery. These works main focus on four types of tasks, including physicochemical property prediction, \textcolor{black}{generative design of crystalline materials}, aiding characterization, and accelerating theoretical computations. Despite the remarkable progress, there is still a lack of systematic investigation to summarize their distinctions and limitations. To fill this gap, we systematically investigated the progress of crystalline materials discovery using artificial intelligence made in recent years. We first introduce several data representations of the crystalline materials. Based on the representations, we summarize various fundamental deep learning models and their tailored usages in various material discovery tasks. Finally, we highlight the remaining challenges and propose future directions. 

\end{abstract}

\begin{document}

\flushbottom
\maketitle

\section{Introduction}


Crystalline materials, possessing unique structures and diverse properties, have been the cornerstone of a wide range of applications such as electronics, sustainable energy, etc~\cite{gomes2021computational,ocp1111dataset,wang2024comprehensive}. They are arranged periodically in the units ---{\em crystal lattice}. Within the crystal lattice, the atoms are arranged meticulously with symmetrical structures, establishing uniform atomic interactions. The long-range order characterized by periodicity and the short-range order featured by symmetry gives rise to their structural stability~\cite{luo2024towardssymmetry,xie2021crystal}. These distinctive structures also induce a diverse array of physical and chemical properties, enabling the versatility of crystalline materials.

The development of new crystalline materials with desired properties is a fundamental problem, which requires a deep understanding of structure-property relationships behind the materials~\cite{vu2023towards,liu2022experimental}. To uncover the structure-property relationships, hitherto, material science has gone through four paradigms~\cite{pyzer2022accelerating}. Before the 17th century, material science primarily relied on empirical and observational methods. In the 17th century, the advent of calculus initiated the second scientific revolution, a transition to theoretical science, characterized by mathematical equations of natural phenomena. The invention of computers in the 20th century created the third paradigm, the computational science paradigm, enabling larger and more complex theoretical equations to become solvable. Through computational simulations, e.g., \textcolor{black}{density functional theory (DFT) calculations}, new materials can be tested and evaluated. However, due to the vast chemical space of compositions and structures, using these simulations to explore the structure-property relationships is of high computational cost and highly depends on the trial-and-error approaches~\cite{pyzer2022accelerating,lan2023adsorbml,xie2021crystal}.

In recent years, the fourth paradigm --- Artificial Intelligence (AI)-powered materials science~\cite{choudhary2020joint,ocp1111dataset}, provides a new avenue for material discovery. Especially, deep learning techniques have revolutionized the field of materials science, enabling the rapid discovery of new materials. One of the key drivers is the availability of large datasets from experiments and simulations~\cite{chanussot2021open,goodall2020predicting}. For example, the Materials Project database includes property calculations for over 60,000 molecules and over 140,000 inorganic compounds of clean energy systems such as photovoltaics, thermoelectric materials, and catalysts~\cite{jain2013commentary}. Additionally, the Open Catalyst Project (OCP) has made available a vast dataset of crystal structures, consisting of 1.3 million structural relaxations with results from over 260 million DFT calculations~\cite{zitnick2020introduction}. More recently, the OMat24 database contains over 110 million DFT calculations, focusing on structural and compositional diversity~\cite{barroso2024open}.

These explosive growths of data have filled the fuel for AI-driven materials science. With these abundant data, deep learning has provided fresh perspectives on various tasks for crystalline material discovery, including physicochemical property prediction, \textcolor{black}{generative design of crystalline materials}, aiding characterization, and accelerating theoretical computations. {\em Physicochemical property prediction} is a fundamental task in deep learning-driven materials science, where the goal is to predict a material's physical and chemical properties. These models typically take the crystal structure as input and output the predicted properties, including scalar properties and tensor properties~\cite{xie2018crystal,schutt2018schnet,chen2019graph,zeni2025generative,cao2024space,schmidt2021crystal}. {\em \textcolor{black}{Generative design of crystalline materials}} is to discover new crystal materials including new compositions and structures. The newly discovered material requires not only stability but also specific properties~\cite{xie2021crystal,zeni2025generative,cao2024space}.
{\em Aiding characterization} is to use deep learning models to aid the quantitative or qualitative analysis of experimental observations and measurements, including assisting in the determination of crystal structure and composition, identifying structural transition, and inferring crystal symmetry and defects~\cite{ziletti2018insightful,yang2021deep,choudhary2022recent,el2019raman,lee2020deep}. {\em Accelerating theoretical computations} aim to reduce the computational cost of intractable simulations by using deep learning models as surrogates to provide potentials, functionals, and electronic structures~\cite{behler2007generalized,chen2022universal,bartok2010gaussian,batzner20223}.

\begin{figure}[h] 
  \centering   
\includegraphics[width=13.7cm]{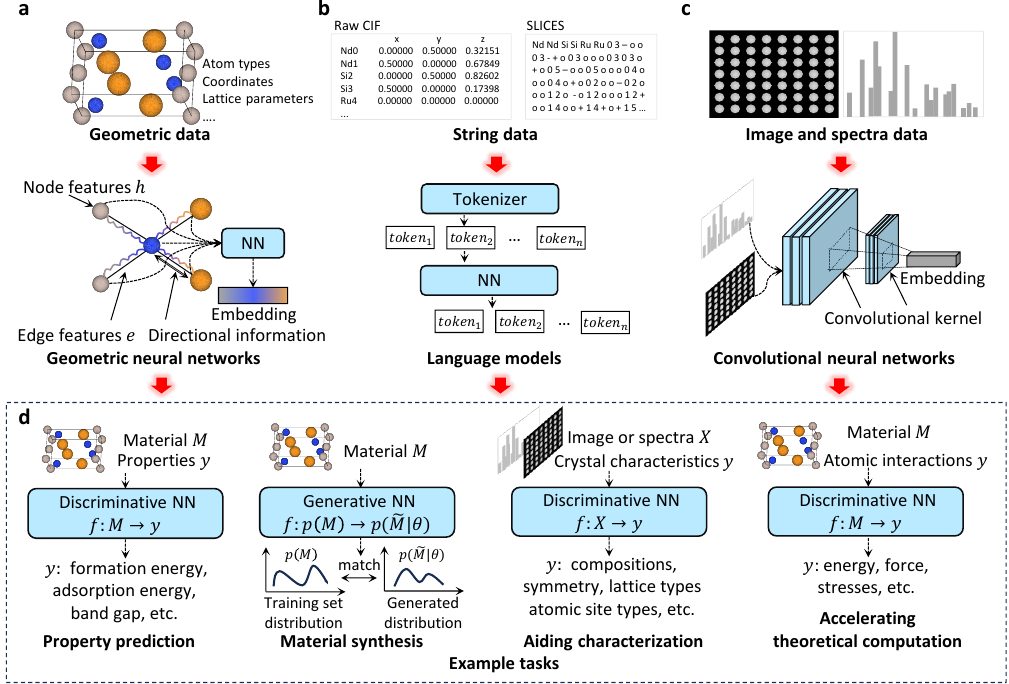} 
  \caption{\textbf{Overview of deep learning for crystalline materials.} \textbf{a|} Crystalline materials are described by geometric graphs. Typically, geometric graph neural networks are used to extract atom, bond, and directional information from the geometric graphs based on message-passing mechanisms. \textbf{b|} Language models are generally used to handle the string representations, such as CIF and SLICES~\cite{xiao2023invertible}. \textbf{c|} The atomic image and spectra data of crystalline materials are processed using convolutional neural networks; the convolutional kernels transform these data into feature map representations. \textbf{d|} The above fundamental models serve as surrogates for various material discovery tasks, including physicochemical property prediction, \textcolor{black}{generative design of crystalline materials}, aiding characterization, and accelerating theoretical computations.
  }
  \label{fig:datamodel}
  \vspace{-0.5cm}
\end{figure}

In this review, we will investigate how AI, especially deep learning techniques, leverages various crystalline material data to learn from crystalline materials for realizing the above four fundamental tasks. 
While several existing reviews have discussed related topics~\cite{choudhary2022recent,han2022geometrically}, they neither cover recently emerged deep learning methodologies for crystalline materials nor have comprehensive coverage of concepts and related tasks to crystalline materials.
Our review distinguishes itself by fulfilling a dual purpose: it functions as both an accessible tutorial introducing fundamental concepts of crystalline materials and state-of-the-art deep learning methodologies, and as a detailed guide to current research developments, highlighting distinctions, advances, and limitations in the domain. Through this dual approach, we aim to provide readers with a broad conceptual overview that bridges AI and materials science, fostering deeper insights into this interdisciplinary field.

In the review, we provide several data representations that have been used in crystalline material research. The data representations involve geometric graphs, string representations, images, and spectra, as shown in FIG. \ref{fig:datamodel}a, b, and c. With different data representations, various fundamental models such as geometric graph neural networks~\cite{reiser2022graph}, language models~\cite{chang2024survey}, and convolutional neural networks~\cite{cong2023review} and their design principles are introduced. Then, we mainly focus on the recently proposed deep learning models on physicochemical property prediction, \textcolor{black}{generative design of crystalline materials}, aiding characterization, and accelerating theoretical computations, as shown in FIG. \ref{fig:datamodel}d. In addition, we provide some of the commonly used datasets, benchmarks, and software used for data-driven crystalline material discovery. It is worth noting that while deep learning has shown its potential in facilitating the development of crystalline material discovery, many challenges and issues need to be addressed to unleash the potential of deep learning for crystalline materials further. Therefore, we also highlight these challenges and provide insights in this domain.
This review aims to offer comprehensive and valuable insights, and fosters progress in the intersection of artificial intelligence and material science. 
We have organized the surveyed work and benchmarking, which can be accessed via the following link: \url{https://github.com/WanyuGroup/AI-for-crystal-materials/}.

\section{Data Representations Describing Crystalline Materials}

Crystalline materials are inherently organized with a periodic arrangement of atoms in the 3D space. The periodic unit, known as the unit cell, is defined by the lattice parameters, which are the lengths of the cell edges and the angles between them. Within the unit cell, an array of atoms is arranged, each of which is defined by its type and atomic coordinates. 
Traditionally, the above information on crystalline materials is documented in text-based formats, namely Crystallographic Information Files (CIFs). CIFs have been the {\em de facto} data representation of crystalline materials~\cite{hall1991crystallographic}, as they record the detailed information about a crystal structure, including atom types, atomic coordinates, lattice parameters, space groups, and other structural attributes as illustrated in FIG. \ref{fig:spectra}a. Several studies simply use CIFs as the string input and employ language models for various downstream tasks~\cite{gruver2024finetuned,cao2024space}. By extracting and interpreting the geometric data from these CIFs, one can generate 3D geometric graphs that accurately represent the overall crystal structures. In addition, researchers can also employ advanced imaging techniques and electromagnetic radiation to obtain atomic images and spectroscopic data. In this regard, this section will provide an overview of commonly used data representations for crystalline materials.

\subsection{Strings}

The simplified line-input crystal-encoding system (SLICES) attempts to provide invertible, invariant, periodicity-aware text-based representations for crystalline materials~\cite{xiao2023invertible}. A quotient graph was proposed to serve as an intermediary to transform between the crystal structures and the SLICES strings~\cite{chung1984nomenclature}. The quotient graph indicates how atoms in a unit cell are connected to atoms within the adjacent unit cells. For instance, as depicted in FIG. \ref{fig:spectra}c, the edge $e_4$ labeled "0 0 1" in the quotient graph, indicates that $e_4$ connects node C$_0$ to the copy of C$_1$ shifted one unit along the $c$ axis. 

SLICES strings convert the quotient graph into three components: atomic symbols, node indices, and edge labels. Specifically, a SLICES string begins with the atomic symbols of the unit cell, encoding the chemical composition of the crystal structure. Edges are explicitly represented in the form $uvxyz$, where $u$ and $v$ are node indices, and $xyz$ denotes the location of the unit cell to connect to. To represent the quotient graph in FIG. \ref{fig:spectra}c, the SLICES string begins with atom symbols C$_0$ and C$_1$ of node indices $0$ and $1$, respectively. Following the atom symbols, the edges $e_1\ldots e_4$ of the form $uvxyz$ in the quotient graph are appended. For example, $e_4$ is represented as "01oo+". Here, $01$ denotes the atomic indices corresponding to the positions in the atomic symbols of the SLICES string (i.e. C$_0$ and C$_1$). The sequence "oo+" represents the label "0 0 1" from the labeled quotient. To reconstruct the original crystal structures from the SLICES strings, Eon's graph theory~\cite{eon2011euclidian} and force field approaches, including geometry frequency noncovalent force field~\cite{spicher2020robust} and M3GNet~\cite{chen2022universal}, can be used to ensure invertibility. On the Materials Project dataset~\cite{jain2013commentary}, the reconstruction routine of SLICES can successfully reconstruct 94.95\% of 40,000 crystal structures which are structurally and chemically diverse, showcasing an unprecedented invertibility.

\subsection{Geometric Graphs}

\begin{figure}[!t] 
  \centering   
  \includegraphics[width=13.7cm]{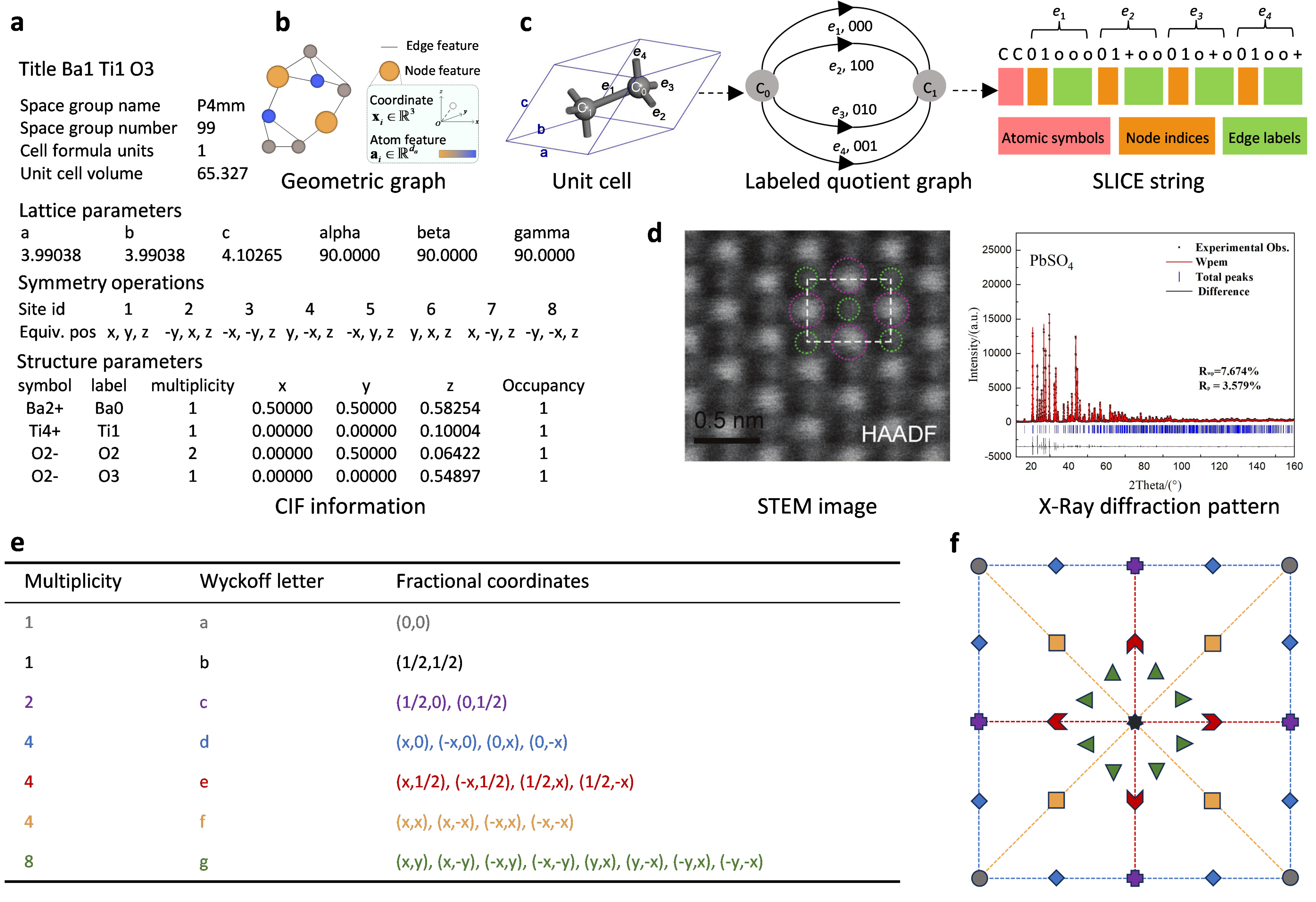} 
  \caption{
\textbf{Examples of data representations: strings, images, and spectra data.}
\textbf{a $|$} The CIF \textcolor{black}{information} of BaTiO$_3$ (MP ID: mp-5986, from the Materials Project~\cite{jain2013commentary}).
\textbf{b $|$} The geometric graph representation.
\textbf{c $|$} The illustration of SLICES string conversion process. 
\textbf{d $|$} The scanning transmission electron microscopy image and the X-ray diffraction spectra data. \textbf{e $|$} The Wyckoff positions of the P4mm space group (No.11).
\textbf{f $|$} The schematic diagram for the Wyckoff positions of the P4mm space group. Panel \textbf{c} adapted from REF.~\cite{xiao2023invertible}. Panel \textbf{d} adapted from REF.~\cite{ge2022atomic} \textcolor{black}{and REF.~\cite{bin2025simxrd}}.
}
  \label{fig:spectra}
  \vspace{-0.8cm}
\end{figure}

With the geometric details of crystalline materials provided in CIFs, it is feasible to model geometric graphs of these materials within the 3D Euclidean space, as shown in FIG. \ref{fig:spectra}b. Next, we introduce two widely utilized coordinate systems, namely the Cartesian coordinate system and the fractional coordinate system. These two coordinate systems are fundamental tools for representing the geometric configuration of crystalline materials~\cite{muller2024symmetry,tilley2020crystals}. Furthermore, based on the corresponding coordinate systems, we will introduce two important concepts in crystal research: the space group and the Wyckoff positions.

\textbf{Cartesian Coordinate System}. A crystalline material can be formally represented as $\mathbf{M}=(\mathbf{A},\mathbf{X},\mathbf{L})$, where $\mathbf{A}=[\boldsymbol{a}_1,\boldsymbol{a}_2,\cdots,\boldsymbol{a}_n]^T\in\mathbb{R}^{n\times d_a}$ denotes the atom feature matrix for $n$ atoms within a unit cell, with $\boldsymbol{a}_{i}\in\mathbb{R}^{d_{a}}$ representing the $d_{a}$-dimensional feature vector of an individual atom, such as the atomic type. The matrix $\mathbf{X}=[\boldsymbol{x}_{1},\boldsymbol{x}_{2},\cdots,\boldsymbol{x}_{n}]^{T}\in\mathbb{R}^{n\times3}$ represents the 3D Cartesian coordinates of $n$ atoms within a unit cell, where $\boldsymbol{x}_i\in\mathbb{R}^3$ specifies the Cartesian coordinates of each atom. The repeated patterns of a crystal material can be described by 
the lattice matrix $\mathbf{L}=[\boldsymbol{l}_{1},\boldsymbol{l}_{2},\boldsymbol{l}_{3}]\in\mathbb{R}^{3\times3}$, where $\boldsymbol{l}_{1}$, $\boldsymbol{l}_{2}$, and $\boldsymbol{l}_{3}$ are a set of basis vectors in the 3D Euclidean space. The unit cell repeats itself along three basis vector directions to form a complete crystal. 
Finally, a complete crystal can be represented as $(\hat{\mathbf{A}}, \hat{\mathbf{X}})=\{(\hat{\boldsymbol{a}_{i}}, \hat{\boldsymbol{x}}_i)|\hat{\boldsymbol{x}}_i=\boldsymbol{x}_i+k_1\boldsymbol{l}_1+k_2\boldsymbol{l}_2+k_3\boldsymbol{l}_3, \hat{\boldsymbol{a}_{i}}=\boldsymbol{a}_{i},k_1,k_2,k_3\in\mathbb{Z},i\in\mathbb{Z},1\leq i\leq n\}$. The integers $k_i$ and $l_i$ represent all possible atomic positions in the periodic lattice.

\textbf{Fractional Coordinate System}. 
The fractional coordinate employs $\mathbf{L}=[\boldsymbol{l}_{1},\boldsymbol{l}_{2},\boldsymbol{l}_{3}]\in\mathbb{R}^{3\times3}$ as the basis vectors for atomic positions. The position of an atom can be described with a fractional coordinate vector \textcolor{black}{$\boldsymbol{f_{i}}=[f_{i,1},f_{i,2},f_{i,3}]^{\top}\in[0,1)^{3}$}. The corresponding Cartesian coordinate vector can be expressed as 
\textcolor{black}{$\boldsymbol{x}_{i}=\sum_j f_{i,j}\boldsymbol{l}_{j}$, where $j=1,2,3$}. Therefore, a crystal $\mathbf{M}$ can be represented as $\mathbf{M} = (\mathbf{A}, \mathbf{F}, \mathbf{L})$, where $\mathbf{F}=[\boldsymbol{f}_1,\cdots,\boldsymbol{f}_n]^{T}\in[0,1)^{n\times 3}$ represents the fractional coordinates of all atoms in the unit cell.

\textbf{\textcolor{black}{Reciprocal Space representations.}} 
\textcolor{black}{
In addition to real space representations such as the Cartesian coordinate system and the fractional coordinate system, crystalline materials can also be described in reciprocal space. Reciprocal space is analogous to the frequency domain of time dependent signals and naturally arises from the 3D Fourier transform of spatial functions. It provides a complementary perspective for characterizing the periodic nature of crystal structures.
Given the lattice matrix of a primitive cell, denoted as $\mathbf{L}_p$, which corresponds to the smallest unit cell by volume, the associated reciprocal lattice vectors are defined as $\mathbf{B} = [\boldsymbol{b}_1, \boldsymbol{b}_2, \boldsymbol{b}_3] = 2\pi (\mathbf{L}_p^{-1})^{T}$, where $\boldsymbol{b}_i$ span the reciprocal lattice. In research on AI for crystal, reciprocal space representations are commonly employed to capture long range interactions that are difficult to model in real space alone~\cite{yu2022capturing,kosmala2023ewald,taniai2024crystalformer,nie2025regnet}.
}

\textbf{Space Group}. 
When we rotate and translate the 3D crystal structures in the 3D space, certain rotation angles and translation distances can map a crystal structure back onto itself due to the inherent symmetry of the structure. These specific rotation and translation transformations are collectively defined as space groups. Specifically, \textcolor{black}{the space group is a discrete subgroup of the Euclidean group $\mathrm{E}(3)$, comprising translations, rotations, inversion, reflections, screw axes operations, and glide planes operations.}
Given a group $g\in\mathrm{E}(3)$ (See Box 1), a crystal $\mathbf{M}$ is recognized to be symmetric with respect to $g$ if $g\cdot\mathbf{M}= \mathbf{M}$ (the symbol $=$ here refers to the equivalence between geometric structures). 
The set of all possible $g$ constitutes the space group of $\mathbf{M}$, and the total number of different space groups is finite. Accordingly, space groups are classified into 230 types for 3D crystal structures~\cite{jiao2024space,hahn1983international}.
The finite space group also imposes strict constraints on specific positions of atoms within the crystal structure, known as Wyckoff positions.

\textbf{Wyckoff Positions}. 
Wyckoff positions describe a set of symmetry-equivalent positions within the unit cell~\cite{janner1983wyckoff,jiao2024space,cao2024space}, which are determined by the space group transformations. Generally, Wyckoff positions include three elements: multiplicity, Wyckoff letter, and fractional coordinates. For instance, FIG. \ref{fig:spectra}e shows the Wyckoff positions for the 2D plain group P4mm \cite{hahn1983international}.
These Wyckoff positions are labeled using Wyckoff letters from the alphabet, which represent distinct symmetry-equivalent sites. The multiplicity indicates the number of these symmetry-equivalent sites. 
\textcolor{black}{To preserve the symmetry of the crystal structure, all these symmetry-equivalent positions within a Wyckoff site are typically occupied by the same atomic species in ordered crystals \footnote{In compositionally disordered crystals, multiple atomic species may occupy the same Wyckoff position with partial occupancies.}.}
\textcolor{black}{FIG. \ref{fig:spectra}e and FIG. \ref{fig:spectra}f present schematic diagrams of the Wyckoff positions}, with symmetry-equivalent positions marked by the same color.
\textcolor{black}{Beyond their definition, Wyckoff positions offer distinct advantages for computational modeling, particularly in crystal structure generation tasks \cite{jiao2024space,levysymmcd,kelviniuswyckoffdiff}. A crystal structure can be fully specified by the atomic species and coordinates within its asymmetric unit, together with the associated space group and Wyckoff positions. All remaining atomic positions in the unit cell are then deterministically generated through symmetry operations. This representation significantly reduces the number of degrees of freedom required to describe a crystal, reducing the likelihood of generating physically invalid or symmetry-breaking structures.
}

\subsection{Images}

Crystalline material image representations are obtained through various advanced imaging techniques, such as diffraction imaging and microscopy~\cite{ziletti2018insightful,yang2021deep,choudhary2022recent}.
Representative diffraction imaging methods include X-ray diffraction (XRD), electron diffraction, and neutron diffraction~\cite{zuo2022data,kardjilov2018advances}.
Microscopy techniques have many different imaging modalities, including scanning transmission electron microscopy (STEM), scanning probe microscopy (SPM), and transmission electron microscopy (TEM)~\cite{bian2021scanning}. For example, STEM images, are shown in FIG. \ref{fig:spectra}d, can reflect the surface morphology and the structure of crystalline materials. These image representations are instrumental in executing specific tasks such as defect detection and classification \cite{ziletti2018insightful, yang2021deep}.

\subsection{Spectra}

The interaction between electromagnetic radiation and materials can generate a measurable spectroscopic signal that varies with the radiation's wavelength or frequency. The spectroscopic signal, generally referred to as {\em spectra data}, is commonly acquired through X-ray diffraction (XRD) and Raman spectroscopy~\cite{el2019raman, lee2020deep, choudhary2022recent}.
As depicted in FIG. \ref{fig:spectra}d, XRD spectra present the plots of X-ray intensity versus the diffraction angle. Therefore, XRD spectra can provide quantitative information about the crystal structure parameters such as peak positions and intensities, which correspond to interplanar spacings and crystallite sizes. On the other hand, the Raman spectra, obtained by measuring the frequency shift of scattered light, can provide characteristic information about crystal structures. This spectral data is instrumental in identifying intrinsic material properties such as chemical composition and crystallographic details~\cite{cong2020application}.

\section{Fundamental Deep Learning Models}
\textcolor{black}{In addition to introducing data representations for describing crystalline materials, we present fundamental deep learning models, including geometric graph neural networks (GGNNs), convolutional neural networks (CNNs), language models, and diffusion models, which have played a cornerstone role in recent advancements. Due to page limits, the related content is provided in the Supplemental Material.}

\begin{table*}[htbp]
  \centering
  \caption{The Summary of Models for Material Property Prediction}
      \rowcolors{2}{gray!25}{white}
\renewcommand\arraystretch{1.0}
  \resizebox{\textwidth}{!}{%
    \begin{tabular}{lllll}
    \toprule
    Methods & Data representations & Fundamental models & Physical knowledge & The predicted properties\\
    \midrule

    SchNet~\cite{schutt2018schnet,schutt2017schnet} &  Geometric graph  &GGNN  &  -   &Formation energy, band gap, etc\\  
    
    CGCNN~\cite{xie2018crystal} &Geometric graph &GGNN & - & Total energy, band gap, etc\\

    MEGNET~\cite{chen2019graph} & Geometric graph  & GGNN &  Global state, e.g. temperature  &Formation energy, band gap, etc\\ 
        
    GATGNN~\cite{louis2020graph} & Geometric graph  &GGNN &   -  &Formation energy, band gap, etc \\ 
        
    ALIGNN~\cite{choudhary2021atomistic,kaundinya2022prediction} & Geometric graph  & GGNN &  -   &Electron DOS, band gap, etc\\

    ECN~\cite{kaba2022equivariant} & Geometric graph &GGNN & Space group  &Formation energy, band gap, etc\\
    
    PotNet~\cite{lin2023efficient} & Geometric graph &GGNN &  Interatomic potentials &Total energy, band gap, etc\\
    CrysGNN~\cite{das2023crysgnn} & Geometric graph &GGNN &  - &Total energy, band gap, etc\\

    ETGNN~\cite{zhong2023general} & Geometric graph & GGNN &   -  &Dielectric, piezoelectric, and elastic tensors\\   
    
DTNet~\cite{mao2024dielectric} & Geometric graph & GGNN &   -  &Dielectric tensor\\ 
    GMTNet~\cite{yan2024space} &  Geometric graph & GGNN &   Space group  &Dielectric, piezoelectric, and elastic tensors\\   

    CEGANN~\cite{banik2023cegann} &  Geometric graph & GGNN &  - & Grain boundary, etc\\
    
    ComFormer~\cite{yan2024complete} &   Geometric graph & Transformer &  -& Total energy, band gap, etc\\        
    Crystalformer (ICLR)~\cite{taniai2024crystalformer} &  Geometric graph & Transformer &   Interatomic potentials &Total energy, band gap, etc\\  
    
\textcolor{black}{CrystalFramer~\cite{ito2025rethinking}} & Geometric graph& Transformer&  Interatomic interactions&Total energy, band gap, etc\\   

    Crystalformer (AAAI)~\cite{wang2024conformal} &  Geometric graph & Transformer  &   - &Total energy, band gap, etc\\ 
    E(3)NN~\cite{chen2021direct} &  Geometric graph&GGNN  &   - & Phonon DOS, electron DOS, etc\\ 
    DOSTransformer~\cite{lee2024density} & Geometric graph & Transformer  &   Energy level of material & Phonon DOS, electron DOS, etc\\ 

    \textcolor{black}{VGNN~\cite{okabe2024virtual}} & Geometric graph & GGNN  &   Momentum dependence & Phonon DOS, $\Gamma$-point phonon, etc\\ 
    
    Matformer~\cite{yan2022periodic} &  Geometric graph & Transformer  &  -& Total energy, band gap, etc\\ 
    CrysDiff~\cite{song2024diffusion} & Geometric graph & Diffusion + GGNN &  Space group  & Total energy, band gap, etc\\
MOFTransformer~\cite{kang2023multi} &  Geometric graph & Transformer &  - & Adsorption properties, band gap, etc\\

    Uni-MOF~\cite{wang2024comprehensive} &  Geometric graph & Transformer &  Global state, e.g. temperature & Adsorption properties\\
    
SODNet~\cite{chenlearning} & Geometric graph & Transformer &   Disordered structure of superconductors  &Critical temperature of superconductors\\

\textcolor{black}{ECSG~\cite{zou2025predicting}} & Geometric graph+Image & GGNN+CNN &  -  &Thermodynamic stability\\

\textcolor{black}{ct-UAEs~\cite{jin2025transformer}} & Geometric graph & Transformer+GGNN &  - &Total energy, band gap, etc\\

\textcolor{black}{AdsMT~\cite{chen2025multi}} & Geometric graph & Transformer &  - &Adsorption properties\\

\textcolor{black}{DPF~\cite{shen2025denoising}} & Geometric graph & Transformer/GGNN &  - &Total energy, band gap, etc\\

\textcolor{black}{PDDFormer~\cite{IJCAI25}} & Geometric graph & Transformer/GGNN &  - &Total energy, band gap, etc\\

\textcolor{black}{Rep-CodeGen~\cite{huangcode}} & Geometric graph & LLM &  - &Total energy, band gap, etc\\

\textcolor{black}{CSLLM~\cite{song2025accurate}} & String & LLM & Crystal symmetry, Wyckoff positions, etc &Synthesizability, etc\\

\textcolor{black}{SciToolAgent~\cite{ding2025scitoolagent}} & String & LLM & Explicit tool dependencies &Adsorption properties, etc\\

\textcolor{black}{SPFrame~\cite{hua2025local}} & Geometric graph & Transformer &  Crystal symmetry &Total energy, band gap, etc\\

    \bottomrule
    \end{tabular}%
    }
  \label{tab:addlabel}%
  \vspace{-0.4cm}
\end{table*}%

\section{Material Discovery Tasks}

\subsection{Physicochemical Property Prediction}

Identifying crystalline materials properties is a complex task that involves theoretical and experimental methods. 
Classical {\em ab initio} physical simulation techniques, such as DFT~\cite{jain2016computational}, can precisely calculate material properties, including electronic structures, phonon spectra, and tensor properties. However, these methods are computationally intensive.
Recently, deep learning techniques offer an efficient way for material properties  predictions~\cite{yan2024complete,taniai2024crystalformer,song2024diffusion,lee2024density,yan2022periodic,lin2023efficient,li2023graph}. These deep neural networks aim to capture the mapping relationship between crystalline materials data and properties from the collected datasets. Current related works can be categorized into GNN-based methods and transformer-based methods. Based on this taxonomy, Table \ref{tab:addlabel} provides a summary of the used data representations, fundamental models, and the predicted properties.

\subsubsection{Geometric Graph Neural Network-Based Prediction}

Due to the inherently periodic and the infinite arrangement of lattice within 3D Euclidean space, GGNN-based property prediction methods necessitate graph construction techniques to effectively represent the infinite lattice and its atomic interactions. Typically, nodes in the graph correspond to atoms and their periodic replicas across the crystal's 3D structure, while edges represent the interactions between these atoms. Following the graph construction, GGNN-based methods employ supervised learning to train the network on a dataset of crystal properties. Once trained, the GGNN is leveraged to predict the properties of crystalline materials.

\begin{figure}[h] 
  \centering   
  \includegraphics[width=13.7cm]{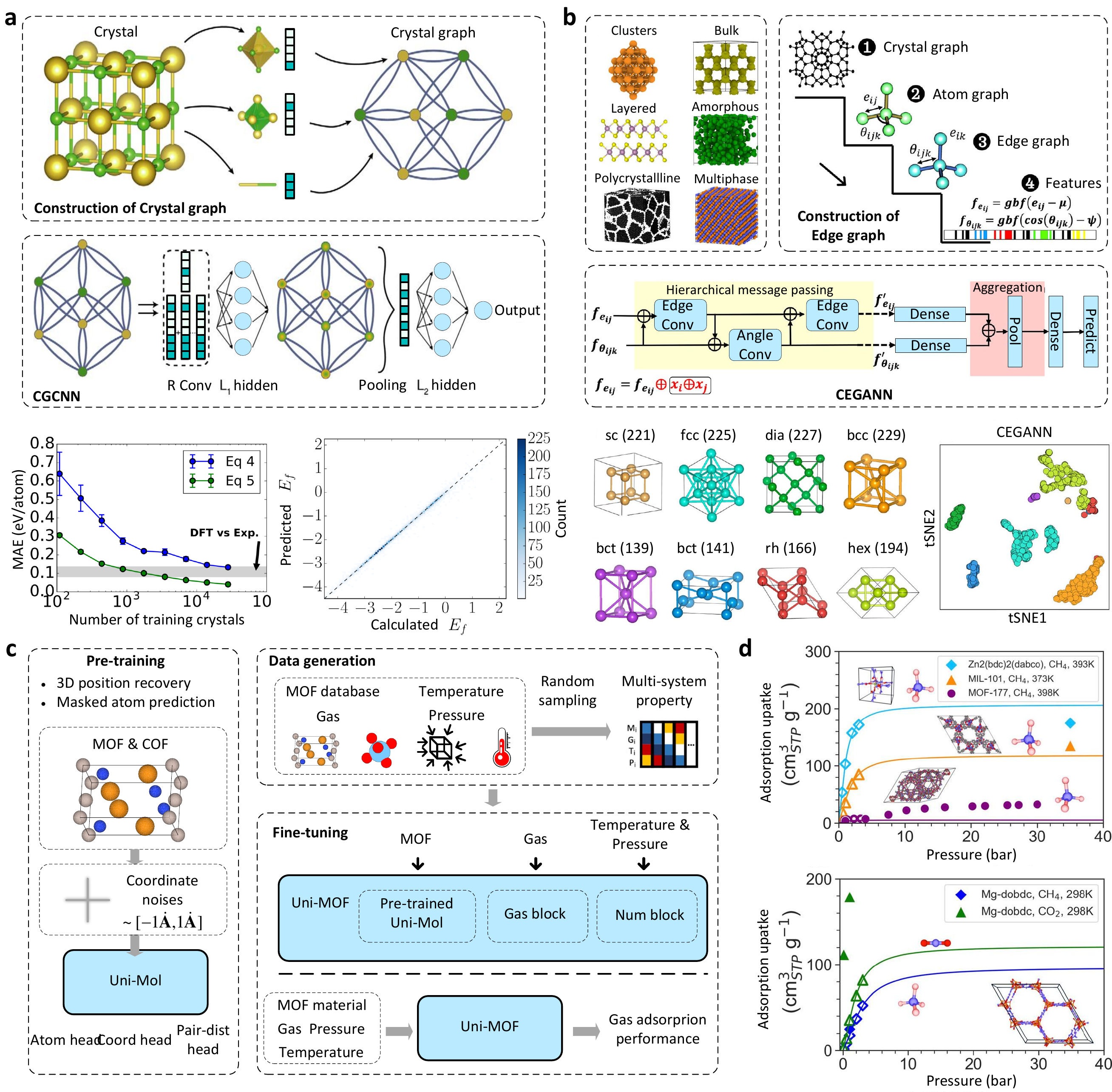}
  \caption{\textbf{GGNN-based and transformer-based methods for property prediction.} 
  \textbf{a $|$} Schematic of CGCNN method. 
  CGCNN experimental results (bottom): \textcolor{black}{The left panel shows the MAE as a function of the number of training crystals for predicting the formation energy using different convolution functions.}
  The right side is a 2D histogram representing the predicted formation energy per atom against the DFT calculated value.
  \textbf{b $|$} Schematic of CEGANN method and the classification experiment results. 
  \textbf{c $|$} Schematic of Uni-MOF method. 
  \textbf{d $|$} 
  Uni-MOF experimental results:
  adsorption isotherms based on low-pressure predictions and high-pressure experimental values.
  Panel \textbf{a} is adapted from REF.~\cite{xie2018crystal}. Panel \textbf{b} is adapted from REF.~\cite{banik2023cegann}. Panel \textbf{c} and \textbf{d} adapted from REF.~\cite{wang2024comprehensive}. 
 }
  \label{fig:overview-property}
  \vspace{-0.5cm}
\end{figure}

An exemplary application of GGNNs is SchNet~\cite{schutt2017schnet,schutt2018schnet}, which was initially designed for simulating quantum interactions in molecules using continuous-filter convolutional layers. This method directly models interactions between atoms utilizing distance information, thereby providing rotation invariant energy predictions.
Although SchNet can effectively handle local environments, it may fall short in capturing more complex long-range interactions and global geometric structures. Compared to SchNet,
CGCNN~\cite{xie2018crystal} is a pioneering GGNN specifically designed for handling crystal structures and it can extract both local and global features. This model was the first to represent crystal structures as multi-edge graphs, as shown in FIG. \ref{fig:overview-property}a. 
The multi-edge connections between nodes reflect the periodicity of the crystal, connecting atoms from different unit cells.
Once the crystal graph is constructed, CGCNN uses convolution and pooling layers to extract and learn both local and global features of the structure. 
As depicted in FIG. \ref{fig:overview-property}a, \textcolor{black}{the bottom left panel reports the performance of CGCNN, where MAE is shown as
a function of training crystals for predicting formation energy per
atom using different convolution functions. The bottom right panel presents a 2D histogram comparing the predicted formation energy per atom with the corresponding DFT-calculated values.} CGCNN has achieved good results. Despite the impressive results, there remains significant potential for further advancements. For instance, CGCNN lacks the capability to account for global state variables (such as temperature), which are essential for predicting state-dependent properties (such as free energy). Additionally, though CGCNN has demonstrated remarkable potential in learning flexible global feature representations, they often underperform in capturing local environments compared to their predictability for global properties like energy band gaps.

Correspondingly, the MEGNet model~\cite{chen2019graph} addresses the first limitation by incorporating global state inputs, including temperature, pressure, and entropy. In MEGNet, the embeddings of global state information are updated simultaneously during the updates of bond and node embeddings. CEGANN~\cite{banik2023cegann} advances the classification potential of GGNNs at both the structural (global) and atomic (local) levels simultaneously. This multi-scale classification method, based on a graph attention architecture, enables the classification of features from atomic-scale crystals to heterogeneous interfaces and micro-scale grain boundaries. As demonstrated in FIG. \ref{fig:overview-property}b, CEGANN begins by constructing an edge graph of the crystal based on the crystal graph. The edge and angle feature representation of the crystal structure is then processed through hierarchical message-passing blocks, generating feature representations for each structure to be used in prediction tasks. This robust approach can be applied to tasks such as the space group classification of crystals (as shown in FIG. \ref{fig:overview-property}b), grain boundary recognition, and the analysis of other heterogeneous interfaces.

GATGNN~\cite{louis2020graph} enhances CGCNN by integrating augmented graph attention layers and global attention layers, thereby leveraging the varying contributions of different atoms in a crystal to the overall material properties. 
The augmented graph attention layer aggregates information from neighboring nodes by calculating attention weights between them, whereas the global attention layer processes the graph's overall structure. 
However, GATGNN also has some drawbacks. For example, the application of the softmax function imposes constraints on the ability of GATGNN to distinguish nodes with different degrees.
Contrasting with traditional GGNN architectures like CGCNN, MEGNet, and GATGNN, ALIGNN employs a line graph neural network derived from the crystal graph~\cite{choudhary2021atomistic,kaundinya2022prediction}. This derived graph describes the connectivity of the edges in the original crystal graph, with its nodes corresponding to atomic bonds and its edges representing bond angles. ALIGNN alternates message passing on these two graphs, leveraging bond lengths and angles in the line graph to incorporate detailed atomic structural information, thus enhancing the model's performance. However, the more intricate graph construction methods employed by ALIGNN result in a heightened computational complexity.

    
Recent advancements in GGNN methodologies have increasingly focused on leveraging the intrinsic properties of crystalline materials \textcolor{black}{or incorporating emerging techniques such as transfer learning to further improve predictive performance.}
Compared to molecules, crystals typically exhibit higher structural symmetry - a property often overlooked by earlier models. Kaba \textit{et al.}~\cite{kaba2022equivariant} addressed this by developing an equivariant model tailored to crystal space groups. 
CrysGNN~\cite{das2023crysgnn} introduces a pre-training framework. During its pre-training phase, CrysGNN reconstructs node features and connectivity in a self-supervised manner. Additionally, it leverages space group and crystal system information to learn structural similarities between graph structures. 
\textcolor{black}{DPF \cite{shen2025denoising} leverages large-scale unlabeled crystal structure data through self-supervised learning. It pretrains a backbone model by reconstructing masked atom types and denoising perturbed atomic positions and lattice parameters, enabling the model to capture intrinsic structural patterns of crystalline materials.
}
PotNet~\cite{lin2023efficient} models interatomic potentials directly as edge features to capture interactions in infinite space. 
This edge feature implicitly encodes information about crystal periodicity, thereby enabling a fully connected GNN to be informed of crystal periodicity.
To address the substantial computational demands of supervised learning approaches for crystal property prediction, Song \textit{et al.}~\cite{song2024diffusion} introduced CrysDiff, a pretrain-finetune framework specifically designed for property prediction tasks.
\textcolor{black}{VGNN \cite{okabe2024virtual} augments GGNNs with carefully designed virtual nodes, thereby overcoming the limitation of fixed output dimensions and enabling the prediction of material-dependent phonon spectra as well as full phonon band structures. ct-UAEs \cite{jin2025transformer} learn task-adaptive and information-rich atomic representations directly from chemical composition and atomic coordinates, without relying on predefined atomic descriptors. The learned embeddings can be integrated into various back-end models, including CGCNN, MEGNET, and ALIGNN, consistently delivering significant accuracy improvements. ECSG~\cite{zou2025predicting} proposes an ensemble learning framework for predicting the thermodynamic stability.}

The previously mentioned GGNNs have primarily focused on predicting single-value properties of crystals. 
These methods cannot be directly applied to tensor property prediction as they fail to achieve $O(3)$ equivariance
for tensor property prediction.
To address this challenge, Edge-based Tensor Prediction Graph Neural Network (ETGNN)~\cite{zhong2023general} represents a crystal's tensor properties as the average tensor contributions from all atoms in the crystal. Each atom's tensor contribution is expressed as a linear combination of local spatial components projected onto the directions of edges in multiple-sized clusters. This tensor decomposition is rotationally equivariant, making ETGNN applicable to tensor property prediction.
Dielectric Tensor Neural Network (DTNet)~\cite{mao2024dielectric} employs a strategy similar to ETGNN, utilizing tensor products to ensure the transfer of tensor equivariance. However, DTNet distinguishes itself by applying tensor products to the equivariant features generated in the network's output, rather than directly on the edge features of the crystal graph. Furthermore, in its experimental evaluation, DTNet was applied to screen the Materials Project dataset, effectively identifying potentially stable materials with high dielectric constants.

Although ETGNN and DTNet achieved good performance, they failed to enforce crystal symmetry constraints in tensor predictions. As an alternative method designed to predict general tensor properties of crystalline materials, GMTNet~\cite{yan2024space} has made improvements in crystal symmetry constraints.
Within the broader GGNN framework, GMTNet adopts the transformer strategy from Comformer~\cite{yan2024complete} to update node-invariant features. It employs spherical harmonic filters and tensor product convolutions to achieve equivariant message passing, updating edge features. Additionally, GMTNet incorporates a crystal symmetry enforcement module that simplifies complex symmetry constraints of tensor properties into constraints applicable to crystal-level features. This architecture enhances the network's robustness against minor errors in message-passing operations and crystal inputs.

\subsubsection{Transformer-Based Prediction}

Recently, transformer-based architectures have shown remarkable capability in graph learning~\cite{ying2021transformers}, and recent works have attempted to apply these architectures to material property prediction. One representative is Matformer \cite{yan2022periodic}, which leverages the geometric distances between atoms from two adjacent unit cells to encode periodic patterns. This enables Matformer to effectively encapsulate the lattice information and periodic patterns. However, Matformer occasionally encodes crystals of different structures as identical graphs \cite{yan2024complete}.
To address this issue, ComFormer \cite{yan2024complete} proposed SE(3) invariant graph representation and SO(3) equivariant graph representation to capture the local and global geometric information by leveraging atomic interatomic distances and atomic angle information. Then, ComFormer converts the two types of graph representations into embeddings by using transformers, including the node-wise transformer layer and edge-wise transformer layer. This enables ComFormer to delicately capture geometric information of different structures.
\textcolor{black}{PDDFormer \cite{IJCAI25} builds upon pairwise distance distribution representations to construct continuous and geometrically complete crystal graphs that remain robust under atomic position perturbations. 
}

On the other hand, Matformer only encodes atomic distance as edge features. To enrich the edge features' expressiveness, CrystalFormer~\cite{wang2024conformal} incorporates angular information into the edge features. In addition, CrystalFormer introduces a graph construction method specifically designed for periodic invariance, using different penalty weights to select appropriate atomic neighborhoods to model the local atomic environments. 
\textcolor{black}{Crystalformer \cite{taniai2024crystalformer} employs infinitely connected attention to model long-range atomic interactions. However, it does not explicitly account for directional information between atoms. CrystalFramer \cite{ito2025rethinking} extends Crystalformer by introducing dynamic local frames that transform interatomic directional information into SE(3)-invariant representations and incorporate them into each message passing layer. SPFrame \cite{hua2025local} further addresses the limitations of local frame based methods, where symmetry equivalent atomic environments may be inadvertently collapsed into indistinguishable representations, leading to broken crystal symmetry. To overcome this issue, SPFrame adopts a local-global associative design that preserves symmetry while maintaining directional expressiveness.}
To further broaden the use of Transformer in property prediction, DOSTransformer~\cite{lee2024density} is specifically designed for the prediction of the density of states (DOS), which is a task not addressed by Matformer. Since DOS is influenced not only by the material itself but also by the energy levels, DOSTransformer takes both the crystal material and energy as heterogeneous input modalities. 
\textcolor{black}{AdsMT \cite{chen2025multi} is a multimodal transformer framework that integrates periodic surface graphs with adsorbate feature vectors through a cross attention mechanism. It directly predicts the global minimum adsorption energy of catalyst surface adsorbate systems without requiring site binding or adsorption configuration information.}

Metal-organic frameworks (MOFs) are a class of crystalline porous materials formed through the self-assembly of metal ions or metal clusters with organic ligands via coordination bonds. Due to their unique porous structure, they have garnered much attention in areas such as gas storage and catalysis. MOFTransformer~\cite{kang2023multi} focuses on MOFs' various properties prediction, including gas adsorption and electronic properties. This work is first pre-trained on extensive MOFs, where graph embedding and energy grid embedding are regarded as multi-modal information to enhance generalizability. Then, the MOFTransformer can be fine-tuned for various properties prediction tasks, e.g., gas adsorption uptake prediction.

MOFTransformer can only predict the adsorption uptake under certain external conditions, which limits the model's generalizability. To make predictions across various external conditions, Uni-MOF~\cite{wang2024comprehensive} is pre-trained on extensive datasets comprising MOFs and covalent organic frameworks (COFs) and subsequently fine-tuned to predict adsorption properties under various conditions such as gas molecules, temperature, and pressure (see FIG. \ref{fig:overview-property}c).
During pre-training, Uni-MOF performs masked atom prediction and introduces noise into the original coordinates of MOFs. This strategy promotes its robustness and generalization capabilities. The fine-tuning stage trains the model under various adsorption conditions, enabling the fine-tuned Uni-MOF to predict adsorption properties under a range of conditions, including different gases, temperatures, and pressures.
As shown in FIG. \ref{fig:overview-property}d, Uni-MOF successfully predicted the gas adsorption isotherms of different MOF structures at various temperatures.

\subsubsection{\textcolor{black}{LLM Meets Crystalline Materials Property Prediction}}
\textcolor{black}{Recent studies have explored the application of LLMs to crystalline materials property prediction. Since LLMs are autoregressive by design, these works typically do not use them to directly perform regression-based property prediction. Instead, LLMs are employed as auxiliary components to support and enhance the prediction process. CSLLM \cite{song2025accurate} utilizes three specialized LLMs to assess the synthesizability of arbitrary 3D crystal structures, recommend plausible synthesis routes, and suggest suitable precursor materials. SciToolAgent \cite{ding2025scitoolagent} introduces a LLM-powered scientific agent that autonomously plans and executes complex multi-tool workflows, enabling informed tool selection through retrieval augmented generation to facilitate property prediction. Rep-CodeGen \cite{huangcode} formulates graph representation learning as a constrained code generation problem and employs multiple LLM agents to iteratively evolve representation code through crossover, evaluation, and selection stages.}
﻿
﻿
﻿
﻿
\subsection{\textcolor{black}{Generative Design of Crystalline Materials}}

\textcolor{black}{Traditionally, materials discovery heavily relied on intuition, trial-and-error experimentation, and serendipity.  However, the chemical space of crystalline structures is immense, making exhaustive exploration through purely trial-and-error experimental approaches time-consuming and resource-intensive. 
In recent years, deep generative models such as GANs, VAEs, diffusion, flow matching, autoregressive generative models, etc, have offered a promising avenue for accelerating crystalline material discovery~\cite{xie2021crystal,jiao2023crystal,cao2024space,zhao2023physics,miller2024flowmm}.
These deep generative models offer a complementary strategy by enabling rapid and large-scale in silico exploration of candidate crystal structures, which can subsequently be validated using traditional experimental and computational techniques.
}
According to data representation, these methods fall into two categories: geometric graph-based generation and string-based generation. Based on this taxonomy, Table \ref{tab:generation} provides a brief summary of the used data representations, fundamental models, and used physical knowledge.

\subsubsection{Geometric Graph-Based Generation}


Due to the bedrock of GGNNs in modeling geometry, a considerable amount of studies use GGNNs to model and generate crystalline materials~\cite{batatia2022mace,schutt2021equivariant,vignac2020building,klipfel2024equivariant}.
One pioneer is G-SchNet~\cite{gebauer2019symmetry}. G-SchNet incorporates the constraints of Euclidean space and the rotational invariances of the atom distribution as prior knowledge. By using the distance between the previously placed positions and the next atomic position as constraints, an equivariant conditional probability distribution is constructed to determine the next atomic position. However, G-SchNet was originally proposed for small molecules, and it falls short of leveraging lattice information to model crystalline materials. 

One representative work specifically designed for crystalline materials is CDVAE~\cite{xie2021crystal}. 
By using the SE(3) invariant GGNN as the encoder, CDVAE explicitly encodes the geometry graph and its lattice as invariant variables. Conditioning on these invariant variables, CDVAE generates materials in a diffusion process that moves atomic coordinates towards a lower energy state and updates atom types to satisfy bonding preferences between neighbors. CDVAE shows nearly 100\% structure validity and composition validity when using the Perov-5 dataset~\cite{castelli2012new} and the MP-20 dataset~\cite{xie2021crystal}. In particular, CDVAE employs a property predictor to guide the diffusion model to generate materials with specific properties. To achieve the goal, CDVAE applies gradient ascent to optimize those invariant variables and uses the property predictor to predict the property of the corresponding generated materials. After optimizing, 90\% generated materials are in the top 15\% of the energy distribution in the MP-20 dataset.

Based on the \textcolor{black}{CDVAE}, researchers have made several extensions. Towards generating stable crystal structure, LCOMs~\cite{qi2023latent} expands on CDVAE to locate the structures with the lowest energy by using gradient-based optimization techniques to the latent variables. After optimizing latent variables, the energy improvement can be up to 5\% on the MatBench dataset~\cite{dunn2020benchmarking}.
\textcolor{black}{Cond-CDVAE}~\cite{luo2024deep} integrates the DimeNet++~\cite{gasteiger2021directional} as the encoder and GemNet-dQ~\cite{gasteiger2021gemnet} as the decoder for conditional generation, making it be capable of generating diverse crystal structures conditioned on various chemical composition and pressure. In particular, it can accurately predict 59.3\% of the 3,547 unseen ambient-pressure experimental structures within 800 structure samplings, with the accuracy rate climbing to 83.2\% for structures comprising fewer than 20 atoms per unit cell.
\textcolor{black}{Con-CDVAE}~\cite{ye2024concdvae} generates crystals' latent variables according to given properties (e.g., formation energy, band gap, crystal system, combination of formation energy and band gap) and then yields the corresponding crystal structure by decoding the latent variables.
\textcolor{black}{MAGECS \cite{park2025exploration} addresses property-driven inverse materials design by coupling deep generative models with efficient global optimization in chemical space. The framework employs CDVAE to generate candidate structures, supervised gated GGNNs to rapidly evaluate target properties, and a bird swarm algorithm to iteratively optimize latent variables toward desired material performance.
}
Towards a generalized material representation, UniMat~\cite{yang2023scalable} defined a 4-dimensional material space that corresponds to the number of periods and groups in the periodic table, the maximum number of atoms per element in the periodic table, and the locations of each atom in a unit cell. The representation shows flexibility for smaller systems, as one can set the number of periods and groups in the periodic table to model specific chemical systems of interest. For example, set the number of periods and groups to 1 for modeling materials containing one specific element. With such a unified representation of materials, UniMat trains diffusion models by treating the representation as a 4-dimensional tensor input or condition. 
On Carbon-24, UniMat outperforms CDVAE in terms of property distribution, composition validity, and structure validity. On the more realistic MP-20 dataset, UniMat achieves the promising composition validity but worse structure validity than CDVAE.

{
\small
\renewcommand\arraystretch{1.0}
\rowcolors{2}{gray!25}{white}

\footnotesize
\setlength{\tabcolsep}{4pt}
\begin{longtable}{p{0.15\textwidth}p{0.25\textwidth}p{0.19\textwidth}p{0.32\textwidth}}
\caption{Summary of Deep Generative Models for Material Generation}
\label{tab:generation} \\

\toprule
\rowcolor{white}
Methods & Data representations & Fundamental models & Physical knowledge/constraints \\
\midrule
\endfirsthead

\toprule
Methods & Data representations & Fundamental models & Physical knowledge/constraints \\
\midrule
\endhead

\midrule
\rowcolor{white}
\multicolumn{4}{r}{\emph{Continued on next page}} \\
\endfoot

\bottomrule
\endlastfoot

     G-SchNet~\cite{gebauer2019symmetry} & Geometric graph: Cartesian sys.  &  GGNN & Non-bonded interaction and  local symmetry\\
    CDVAE~\cite{xie2021crystal} & Geometric graph: Cartesian sys. & VAE + Diffusion + GGNN & Harmonic force field approximation \\
Con-CDVAE\cite{ye2024concdvae} & Geometric graph: Cartesian sys. & VAE + Diffusion + GGNN & Same as CDVAE; property constraints \\
Cond-CDVAE\cite{luo2024deep} & Geometric graph: Cartesian sys. & Diffusion & Same as \textcolor{black}{CDVAE}; property constraints \\
     LCOMs~\cite{qi2023latent}& Geometric graph: Cartesian sys.  & Diffusion & Same as CDVAE; property constraints  \\ 
         DiffCSP~\cite{jiao2023crystal} & Geometric graph: fractional sys. & Diffusion + GGNN & Periodic equivariance \\
DiffCSP-SC~\cite{chenlearning} & Geometric graph: fractional sys. & Diffusion + GGNN & Periodic equivariance \\
         
EquiCSP~\cite{linequivariant}& Geometric graph: fractional sys. & Diffusion + GGNN & Lattice equivariance \\
         
GemsDiff~\cite{Klipfel_Fregier_Sayede_Bouraoui_2024}& Geometric graph: Cartesian sys.  & Diffusion + GGNN &  Composition constraint\\ 
       SyMat~\cite{luo2024towardssymmetry} & Geometric graph: Cartesian sys. & VAE + Diffusion + GGNN &  Permutation and rotation invariance \\
            EMPNN~\cite{klipfel2024equivariant}& Geometric graph: Cartesian sys.  & GGNN &  Lattice equivariance \\ 
            UniMat~\cite{yang2023scalable} &  
Geometric graph: Cartesian sys. &  Diffusion & -
\\    MatterGen~\cite{zeni2025generative}& Geometric graph: fractional sys.  & Diffusion + GGNN & -  \\ 
    PGCGM~\cite{zhao2023physics}& Geometric graph: fractional sys.  & GAN &  Crystal symmetry constraint\\ 
CubicGAN~\cite{zhao2021highcubiccrystal}& Geometric graph: fractional sys.  & GAN &  Symmetry constraint\\ 
    PCVAE~\cite{10191051}& Geometric graph: Cartesian sys.  & VAE &  Lattice and symmetry constraint\\ 
       DiffCSP++~\cite{jiao2024space} & Geometric graph: fractional sys. & Diffusion + GGNN & Crystal symmetry constraint\\
       FlowMM\cite{miller2024flowmm}& Geometric graph: fractional sys. & Flow Matching + GGNN & Crystal symmetry constraint \\
Govindarajan~\cite{govindarajan2023behavioral,govindarajan2023learning}  & Geometric graph: Cartesian sys.  & GGNN   & - \\
   CHGFlowNet ~\cite{nguyen2023hierarchical}  & Geometric graph: Cartesian sys.  & GGNN  & Symmetry constraint  \\
         LM-CM,LM-AC~\cite{flam2023language} & String: CIFs  & LLM &  - \\ 
         CrystaLLM~\cite{antunes2024crystal} & String: CIFs  & LLM  &  - \\ 
    CrystalFormer~\cite{cao2024space}& String: CIFs  & \textcolor{black}{Autoregressive Transformer} & Crystal symmetry constraint \\ 
    SLI2Cry~\cite{xiao2023invertible} & String: SLICES & RNN   &  - \\
Gruver~\cite{gruver2024finetuned} &  String: CIFs & LLM & -\\

FlowLLM~\cite{sriramflowllm} &  Geometric graph: Cartesian sys.& LLM + Flow Matching  & -\\

Mat2Seq~\cite{Mat2Seqyan} &  String & LLM  & Crystal symmetry constraint\\
GenMS~\cite{yang2024generative} &  String: CIFs & LLM + Diffusion  & Property constraints\\

ChemReasoner~\cite{sprueill2024chemreasoner} &  String & LLM + GGNN  & Property constraints\\

\textcolor{black}{ FlowDPO~\cite{jiao20243d}} &  Geometric graph: fractional sys. & Flow Matching+GGNN &-\\

\textcolor{black}{ a$^2$c~\cite{aykol2025predicting} } &  Geometric graph: fractional sys. & GGNN &-\\

\textcolor{black}{ CrystalMath~\cite{galanakis2024rapid} } & -& General machine learning &Crystal symmetry, packing topology, etc.\\

\textcolor{black}{ SymmCD~\cite{levysymmcd} } & Geometric graph: fractional sys.& Diffusion + GGNN &Crystal symmetry constraint\\

\textcolor{black}{ MatExpert~\cite{dingmatexpert} } & String & LLM &Expert-inspired retrieval\\

\textcolor{black}{ MOFFlow~\cite{kimmofflow} } & Geometric graph: Cartesian sys. & Flow Matching+GGNN &-\\

\textcolor{black}{ TGDMat~\cite{dasperiodic} } & Geometric graph: fractional sys. & Diffusion + GGNN &Text descriptions\\

\textcolor{black}{ CrysBFN~\cite{wu2025periodic} } & Geometric graph: fractional sys. & Bayesian Flow + GGNN &-\\

\textcolor{black}{ OSDA~\cite{hu2025osda} } & - & LLM &Chemical rules, binding-energy evaluation, etc\\

\textcolor{black}{ Shaul~\textit{et al.}~\cite{shaulflow} } & - & Flow Matching &kinetic-energy optimality\\

\textcolor{black}{ MAGUS~\cite{han2025efficient} } & Geometric graph: Cartesian sys. & General machine learning  &Crystal symmetry constraint \\

\textcolor{black}{ Target XXXI~\cite{zhou2025robust} } & - & General machine learning  &Crystal symmetry constraint \\

\textcolor{black}{ Chemeleon~\cite{park2025exploration} } & Geometric graph: fractional sys.  & Diffusion + GGNN  &Text descriptions \\

\textcolor{black}{ MAGECS~\cite{park2025exploration} } & Geometric graph: Cartesian sys.  & Diffusion + GGNN  &Property-guided global search in chemical space \\

\textcolor{black}{ WyFormer~\cite{kazeevwyckoff} } & String & Autoregressive Transformer  &Crystal symmetry constraint\\

\textcolor{black}{ KLDM~\cite{cornet2025kinetic} } & Geometric graph: fractional sys. & Diffusion + GGNN   &-\\

\textcolor{black}{ WyckoffDiff~\cite{kelviniuswyckoffdiff} } & Geometric graph: Cartesian sys. & Diffusion + GGNN   &Crystal symmetry constraint\\

\textcolor{black}{ OMatG~\cite{hollmer2025open} } & Geometric graph: fractional sys. & Diffusion/Flow Matching + GGNN   &-\\

\textcolor{black}{ ADiT~\cite{joshiall} } & Geometric graph: fractional sys. & Diffusion + Transformer   &-\\

\textcolor{black}{ MACS~\cite{zamaraeva2025macs} } & Geometric graph: Cartesian sys. & Multi-agent reinforcement learning  &-\\

\textcolor{black}{ SGEquiDiff~\cite{chang2025space} } & Geometric graph: fractional sys. & Diffusion + GGNN + Autoregressive Transformer   &Crystal symmetry constraint\\

\textcolor{black}{ CrysLLMGen~\cite{khastagir2025llm} } & String+Geometric graph: fractional sys. & LLM + Diffusion + GGNN   &-\\

\textcolor{black}{ MOF-BFN~\cite{jiao2025mof} } & Geometric graph: fractional sys. & Bayesian Flow+ GGNN  &-\\

\textcolor{black}{ MOFFlow-2~\cite{kim2025flexible} } & String + Geometric graph: fractional sys. & Flow Matching + GGNN + Autoregressive Transformer   &-\\
\textcolor{black}{ CrystalICL~\cite{wang2025crystalicl} } & String & LLM   &Crystal symmetry constraint\\

\end{longtable}
}

\begin{figure}
  \centering   
  \includegraphics[width=13.7cm]{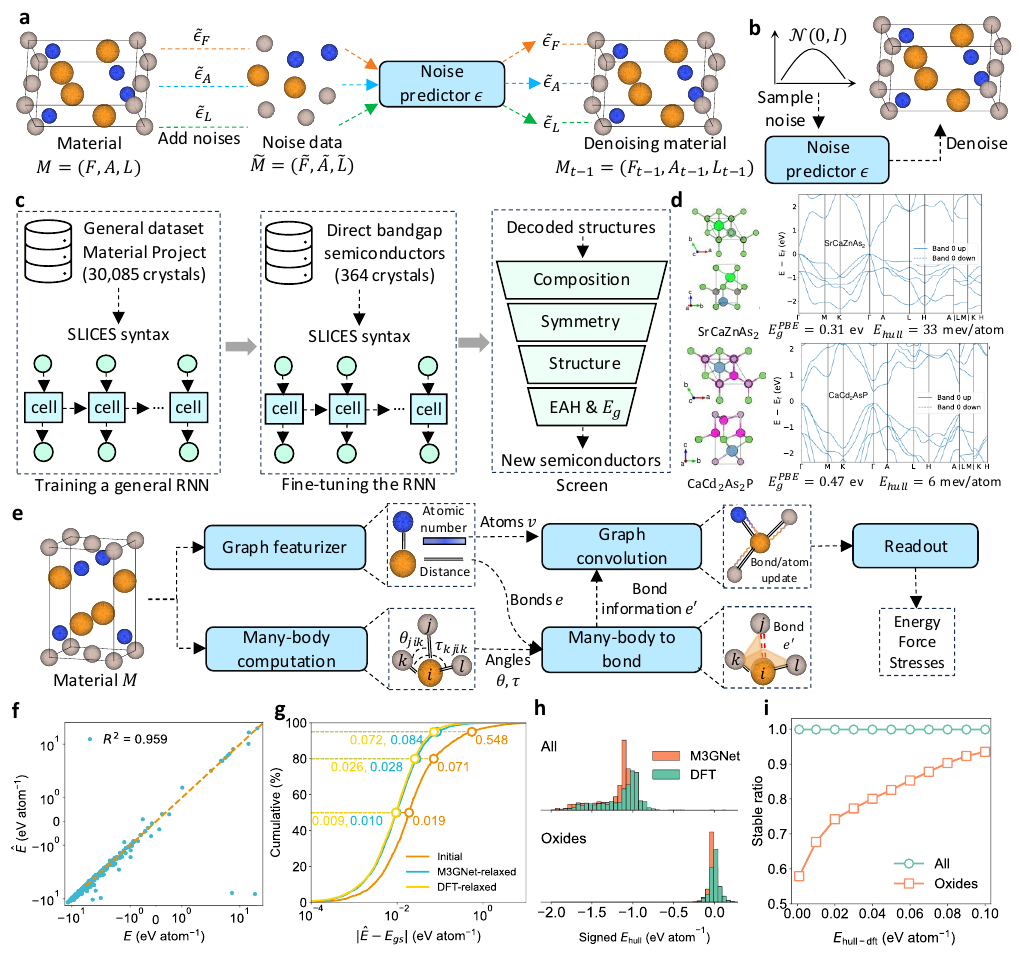}
\vspace{-0.4cm}
  \caption{\textbf{GGNN-based and string-based methods for materials generation.} \textbf{a $|$} For the training stage, DiffCSP adds noises and denoises for atoms, factorial coordinates, and lattice parameters. \textbf{b $|$} For the generation stage, random noises are sampled from Gaussian space for denoising as new materials. \textbf{c $|$} Schematic overview of SLI2Cry. The CIF data are converted to SLICES strings and then used to train a general RNN. The RNN is then fine-tuned on a direct bandgap semiconductors dataset. A number of SLICES strings are sampled from the RNN and decoded, then new direct narrow-gap semiconductors can be filtered. \textbf{d $|$} The generated materials with their PBE band structures, the values of bandgap at the PBE level ($E_{g}^{PBE}$), and energy above hull ($E_{hull}$). 
  \textbf{e $|$} In M3GNet, the graph featurizer extracts the atom and bond distance information. Many-body computation calculates the angles $\theta$ and $\tau$ between three bodies and many bodies, respectively. Many-body to bond calculates the new bond information by considering the full bonding environment via angles $\theta$, $\tau$, and bond distance. The readout obtains the final embedding for predicting energy, force, and stresses. \textbf{f $|$} $R^2$ values for the linear fitting between DFT calculated energy and the predicted energy. \textbf{g $|$} The differences between M3GNet-predicted energies $\hat{E}$ and ground state energies $E_{gs}$ using the initial, M3GNet-relaxed, and DFT-relaxed structures. $E_{gs}$ is defined as the DFT energy of the DFT-relaxed crystal. The horizontal lines mark the 50th, 80th, and 95th percentiles of the distributions, and the corresponding x-axis values are annotated. \textbf{h $|$} The signed $E_{hull}$ distribution for the top-1000 lowest $E_{hull-m}$ materials from any chemistry and oxides only. \textbf{i $|$} Fraction of materials below $E_{hull-dft}$ among top-1000 materials in the All and Oxides categories. Panels \textbf{a} and \textbf{b} adapted from REF.~\cite{jiao2024space}. Panels \textbf{c} and \textbf{d} adapted from REF.~\cite{xiao2023invertible}. Panels \textbf{e}$\sim$\textbf{i} adapted from REF.~\cite{chen2022universal}.
  } 
  \label{fig:overview-generation}
\end{figure}

Early works often used Cartesian coordinate systems to represent and model geometric graphs. Recently, a number of studies have used the fractional coordinate system instead of Cartesian coordinate systems to better reflect the symmetry of the underlying lattice. DiffCSP~\cite{NEURIPS2023_38b787fc} utilizes the fractional coordinate system to intrinsically represent crystals and model periodicity. In particular, by employing an equivariant graph neural network for the denoising process, DiffCSP~\cite{NEURIPS2023_38b787fc} introduced an equivariant diffusion approach, which conducts joint diffusion on lattices and fractional coordinates to comprehensively capture the crystal geometry, thereby enhancing the modeling of the crystal geometry. 
As shown in FIG. \ref{fig:overview-generation}a, DiffCSP separately and simultaneously adds noise process and denoises on the fractional coordinates $F$, atom types $A$, and lattices $L$.
For the generation stage, random noises are sampled from Gaussian space for denoising as new materials, as shown in FIG. \ref{fig:overview-generation}b.
DiffCSP achieves comparable validity with previous methods, e.g., CDVAE, but significantly outperforms previous methods in terms of the target properties.
Following this, Chen~\textit{et al.}~\cite{chenlearning} extended DiffCSP for superconductor generation. In addition, EquiCSP \cite{linequivariant} keeps lattice permutation equivariance in diffusion models by incorporating a permutation invariance penalty term during the denoising model training. EquiCSP maps all equivalent periodic data to the same representation and addresses periodic translation invariance. Compared to the DiffCSP method, EquiCSP fully realizes E(3) equivariance based on periodic graph symmetry during the diffusion training process.
\textcolor{black}{FlowDPO \cite{jiao20243d} adopts the denoising network of DiffCSP as a backbone and trains flow-based generative models using Direct Preference Optimization, combining flow matching with automatically constructed preference datasets to guide generation toward structurally consistent outcomes.
CrysBFN \cite{wu2025periodic} introduces a periodic Bayesian flow defined on a hyper torus to explicitly model fractional atomic coordinates, together with entropy conditioned Bayesian updates that better capture uncertainty during generation. TGDMat \cite{dasperiodic} injects global structural knowledge from textual prompts at each denoising step via a pre-trained MatSciBERT \cite{gupta2022matscibert} model, enabling controllable generation that aligns with user-specified constraints. Chemeleon \cite{park2025exploration} integrates cross-modal contrastive learning to align textual descriptions with equivariant crystal graph embeddings and employs a classifier-free denoising diffusion model for text-guided crystal generation.
}
\textcolor{black}{Alternative generative paradigms have also been explored to improve scalability and efficiency. ADiT \cite{joshiall} embeds all atom representations, including atom types, coordinates, and lattice parameters, into a shared latent space using a variational autoencoder and performs generation within this space via a Diffusion Transformer with classifier free guidance. By avoiding explicitly equivariant architectures, ADiT achieves state of the art generation quality while significantly improving scalability and inference efficiency. OMatG \cite{hollmer2025open} proposes a unified generative framework for inorganic crystal discovery based on stochastic interpolants, which generalize both score based diffusion models and conditional flow matching. From a theoretical perspective, Shaul et al. \cite{shaulflow} expand the design space of discrete generative models by formulating flow matching over general discrete probability paths from a kinetic-optimal perspective. KLDM \cite{cornet2025kinetic} explicitly models fractional atomic coordinates on a hypertorus using kinetic Langevin dynamics, coupling atomic positions with auxiliary Euclidean velocity variables to enable diffusion in a flat space while rigorously preserving periodic translation symmetry.
}

To improve the generalizability across the periodic table, Zeni~\textit{et al.}~\cite{zeni2025generative} presented MatterGen, a model that generates stable, diverse inorganic materials across the periodic table. To enable this, MatterGen uses a diffusion model to produce crystalline structures by gradually refining atom types, coordinates, and the periodic lattice. In particular, an equivariant score network is pre-trained on a large dataset of stable material structures to jointly denoise atom types, coordinates, and the lattice. The score network is then fine-tuned with a labeled dataset, where the property labels are encoded to steer the generation towards a broad range of property constraints, such as desired chemistry, symmetry, and scalar property constraints.
\textcolor{black}{In contrast to earlier approaches such as CDVAE and DiffCSP, which primarily emphasize basic structural and compositional validity, MatterGen explicitly targets practical applicability by prioritizing the generation of \emph{stable} materials, followed by material diversity. To this end, MatterGen~\cite{zeni2025generative} introduced the \emph{Stability, Uniqueness, and Novelty (S.U.N.)} metrics for model evaluation, which have since been widely adopted in subsequent studies. Empirically, compared to prior generative models such as \textcolor{black}{CDVAE} and G-SchNet, structures generated by MatterGen are more than twice as likely to be both novel and stable, and are over fifteen times closer to local energy minima.}



\textcolor{black}{Some researchers have begun to focus on application specific tasks such as metal organic framework (MOF) design. MOFFlow \cite{kimmofflow} targets scalable and accurate structure prediction of MOFs by leveraging flow matching on Riemannian manifolds. MOFFlow-2 \cite{kim2025flexible}, a follow up to MOFFlow, presents a two stage generative framework for MOF design and structure prediction that jointly models chemical and geometric degrees of freedom. In the first stage, an autoregressive Transformer generates novel metal clusters and organic linkers as canonicalized SMILES sequences, which are then initialized into 3D building blocks using cheminformatics tools. In the second stage, a torsion aware flow matching model assembles these building blocks by jointly predicting translations, rotations, lattice parameters, and torsion angles on appropriate manifolds. MOF-BFN \cite{jiao2025mof} further advances this direction by decomposing MOFs into rigid building blocks and jointly modeling lattice parameters, fractional block positions, and block orientations within a unified Bayesian updating framework.
}

Recently, several studies also sought to enhance the performance of generative models by integrating prior physical knowledge, such as introducing space group constraints directly into the model.
Zhao~\textit{et al.} proposed a GAN model, called CubicGAN~\cite{zhao2021highcubiccrystal}, to generate large-scale cubic materials conditioning on the atom types and space groups. Further, Zhao \textit{et al.}~\cite{zhao2023physics} also designed PGCGM, a physics-guided deep learning model for generating crystalline materials with high symmetry. Particularly, PGCGM devises two physics-oriented losses: atomic pairwise distance loss and structural symmetry loss.
By using DFT calculations, 1869 out of 2000 generated materials were successfully optimized, of which 39.6\% have negative formation energy, and 5.3\% have energy-above-hull less than 0.25 eV/atom, indicating their thermodynamic stability and potential synthesizability. However, the application of PGCGM is constrained by ternary systems, thus limiting its universality. 
SyMat~\cite{luo2024towardssymmetry} aims to capture physical symmetries of periodic material structures, in which a score-based diffusion model generates atom coordinates and calculates coordinate matrix as scores to ensure symmetry. Compared to CDVAE, the rate of generated materials in the top 15\% of energy distribution in the MP-20 dataset has increased by 7\%.

﻿
\textcolor{black}{A line of work directly constructs specialized data representations \cite{zhu2024wycryst}, such as Wyckoff template-based generation, to guarantee that generated samples strictly satisfy space group constraints. 
Owing to crystal symmetry, lattice matrix elements in different space groups are subject to specific structural constraints. DiffCSP++ \cite{jiao2023space} generates only the unconstrained lattice elements, thereby simplifying the generation process. It further reconstructs atomic fractional coordinates and element types by deriving symmetry equivalent atoms from a single representative atom through symmetry operations, ensuring that the generated crystals fully comply with space group constraints. SymmCD \cite{levysymmcd} decomposes a crystal into an asymmetric unit and its associated symmetry transformations, and jointly models their distributions using a combination of continuous and discrete diffusion processes. Unlike DiffCSP++, which generates full lattice matrices based on predefined templates, SymmCD generates only the asymmetric unit along with its unit parameters. WyFormer \cite{kazeevwyckoff} is an autoregressive generative approach that, for symmetry related atoms, predicts only discrete attributes such as space group, element type, site symmetry, and enumeration. The complete crystal structure is subsequently reconstructed by combining these discrete attributes with energy relaxation. WyckoffDiff \cite{kelviniuswyckoffdiff} performs diffusion directly in a discrete Wyckoff based representation rather than over continuous atomic coordinates. It encodes crystal structures as protostructures defined by space groups and Wyckoff position occupancies, and applies a discrete denoising diffusion model with a graph neural network backbone to model symmetry consistent transitions. SGEquiDiff \cite{chang2025space} further factorizes crystal generation into space group selection, symmetry constrained lattice sampling, autoregressive prediction of Wyckoff positions and elements, and space group equivariant diffusion of atomic coordinates using a Space Group Wrapped Normal distribution. By theoretically showing that equivariant vector fields remain within the tangent spaces of Wyckoff positions, SGEquiDiff avoids post hoc projection and yields space group invariant likelihoods. Compared to DiffCSP++, SGEquiDiff achieved a higher S.U.N. rate.}




In addition, a handful of works explore using reinforcement learning methods to autoregressively generate materials. Govindarajan~\cite{govindarajan2023behavioral,govindarajan2023learning} formulates the problem of designing new crystals as a sequential prediction task. At each step, given an incomplete graph of a crystal skeleton, an agent assigns an element to a specific node. The policy network is a graph neural network that transforms a given state into an effective representation and predicts the action. Similarly, CHGFlowNet ~\cite{nguyen2023hierarchical} used a graph neural network-based hierarchical policy network to generate materials, where high-level decision-making policy operates on the space groups, and the low-level execution policy operates on the atom-lattices policy actions. These reinforcement learning methods show comparable performance and provide a new avenue for the exploration of chemical space.

\subsubsection{String-Based Generation}

Crystalline materials are stored in standard text file formats known as CIFs. By treating CIFs as plain string representations, massive works explore the use of generative language models to generate crystals. The recently proposed invertible and invariant crystal representation, SLICES, also provides a new avenue for using language models in material generation. In this part, we review these string-based generative models.

Several works simply use CIFs as inputs. Flam-Shepherd \textit{et al.}~\cite{flam2023language} uses sequences of discrete tokens to represent everything in CIFs, such as atom coordinates and atom types. By encoding the data as tokens, it is simple to adapt LLMs to many different kinds of molecular structures, including small molecules, protein-binding pockets, and crystals. The work demonstrates that language models trained from scratch on many common molecular datasets actually outperform popular domain-specific models in their ability to generate valid compositions and structures. On the MP-20 dataset, LM-CM and LM-AC show nearly 90\% structure validity and 90\% composition validity.
Also, Antunes \textit{et al.}~\cite{antunes2024crystal} developed a language model called CrystaLLM to generate crystal structures as discrete sequences by training from scratch on millions of CIF strings. Particularly, the integration with an energy predictor permits the use of a Monte Carlo tree search algorithm to search the structures with lower formation energy. 
\textcolor{black}{MatExpert~\cite{dingmatexpert} focuses on crystal material generation by mimicking the multi-step reasoning process of human materials experts using LLMs. The framework decomposes materials discovery into three stages, namely retrieval, transition, and generation, where it first retrieves a reference material with similar properties, then reasons about modification pathways, and finally generates a new crystal structure in CIF format. CrystalICL~\cite{wang2025crystalicl} proposes the first crystal generation framework explicitly designed to leverage the in-context learning (ICL) capabilities of LLMs. The method introduces a space-group-based crystal tokenization scheme that compresses 3D crystal structures into symmetry-aware textual representations using Wyckoff positions, significantly reducing the burden of learning crystal symmetry.
}

Cao \textit{et al.}~\cite{cao2024space} introduced CrystalFormer, a transformer-based autoregressive model for space group-controlled crystalline materials generation. In CrystalFormer, the crystalline materials are represented by the Wyckoff letter, chemical elements, and fractional coordinates. To generate new samples, the transformer sequentially samples atom types, coordinates, Wyckoff positions, and lattice parameters. 
Gruver \textit{et al.}~\cite{gruver2024finetuned} showed that fine-tuned LLMs can generate the three-dimensional structure of stable crystals as text. The work converts the crystal lattice, atom identities, and atom positions into strings, and a LLM, LLaMA-2, is fine-tuned on the strings of inorganic materials. By using string formatted crystals and task-specific prompting, the work enables unconditional stable materials generation, text-condition materials generation, and structural infilling. By using DFT calculations, Gruver \textit{et al.} showed that the fine-tuned LLaMA-2 can generate materials predicted to be metastable at about twice the rate (49\% vs 28\%) of CDVAE.

Due to the direct utilization of CIF files or similar text representations without a special process, these previously mentioned string-based generation methods often fail to ensure SE(3) invariance and periodic invariance within the model. To address this limitation, Mat2Seq~\cite{Mat2Seqyan} converts 3D crystal structures into 1D sequences through techniques such as Niggli cell reduction and irreducible atom sets. This approach ensures that varied mathematical descriptions of an identical crystal structure are consolidated into a single unique sequence, thereby achieving SE(3) and periodic invariance. Another way to ensure SE(3) invariance and periodic invariance is to generate string representation that satisfies both invertibility and invariances such as SLICES.
SLI2Cry\textit{et al.}~\cite{xiao2023invertible} applies SLICES for the inverse design of direct narrow-gap semiconductors for optoelectronic applications. The schematic overview of SLI2Cry is given in FIG. \ref{fig:overview-generation}c. Firstly, a general recurrent neural network (RNN) is trained on the Materials Project database to learn the syntax of SLICES strings and fine-tuned on a dataset of direct narrow-gap semiconductors. The fine-tuned RNN was used to generate large volumes of SLICES strings. Then, a great number of SLICES strings are sampled from the RNN and decoded into crystal structures. By filtering composition, symmetry, structure, and energy above hull (EAH) and $E_g$, 14 new direct narrow-gap semiconductors are identified. In FIG. \ref{fig:overview-generation}d, examples of new materials and their Perdew-Burke-Ernzerhof (PBE) band structures are provided.

Beyond the direct use of LLMs to generate crystal representations, recent approaches have explored integrating LLMs with other generative techniques to enhance model performance in crystal structure generation. For example, FlowLLM~\cite{sriramflowllm} combines LLMs with Riemann Flow Matching (RFM) to improve generation fidelity. In this approach, the LLM is first fine-tuned to learn the data distribution of metastable crystals, enabling it to generate text representations of these structures. These string representations are then converted into geometric graph representations, and then RFM iteratively refines the atom coordinates and lattice parameters to produce stable crystal structures. FlowLLM achieves a roughly 50\% improvement in stable, unique, and novel generation rates compared to previous models.

GenMS~\cite{yang2024generative} integrates LLMs with diffusion models. Initially, GenMS utilizes the LLM to generate intermediate textual representations of the crystal, such as chemical formulas. This intermediate information is then input into a diffusion model to produce crystal structures. GenMS then incorporates a neural network to predict the properties of the generated crystals, allowing for effective screening and selection of desirable structures. 
\textcolor{black}{CrysLLMGen \cite{khastagir2025llm}, similar to GenMS, integrates LLMs with diffusion models. In contrast to GenMS, CrysLLMGen first employs a fine-tuned LLM to produce an intermediate representation consisting of atom types, atomic coordinates, and lattice parameters, leveraging the strong capability of LLMs in modeling discrete compositions. The predicted atomic types are then fixed, while a diffusion model, which is more effective for continuous variables, is used to refine the atomic coordinates and lattice structure. By combining the complementary strengths of LLMs and diffusion models, CrysLLMGen achieves substantial performance improvements. Compared to diffusion only approaches such as SymmCD, CrysLLMGen achieves relative gains of 36.27\% in S.U.N. rate.
}

ChemReasoner~\cite{sprueill2024chemreasoner} is a LLM-driven heuristic search framework for catalyst discovery. This work frames catalyst discovery as a search problem, where an agent, e.g., GPT-4, is informed by computational chemistry feedback to search favorable catalysts. Specifically, catalysts identified in the intermediate search steps are evaluated structurally based on spatial orientation, reaction pathways, and stability. A scoring function, which considers adsorption energies and reaction energy barriers, directs the LLM to focus on energy-efficient and effective catalysts.
\textcolor{black}{OSDA~\cite{hu2025osda} employs an Actor-Evaluator-Self-reflector loop, where an LLM generates candidate OSDA molecules, chemical tools evaluate their validity, synthesizability, and binding energy, and reflective feedback iteratively refines subsequent generations. 
}

\subsubsection{\textcolor{black}{Other Paradigms}}

\textcolor{black}{Beyond string-based and geometric graph-based generation, several alternative paradigms have been explored for crystal design. a$^2$c \cite{aykol2025predicting} targets crystallization from amorphous precursors by sampling local structural motifs from atomistic amorphous models and relaxing extracted subcells under periodic-boundary conditions using universal interatomic potentials, enabling the discovery of metastable and stable crystalline phases. CrystalMath \cite{galanakis2024rapid} adopts a purely mathematical and topological approach for molecular crystal structure prediction, formulating analytically defined objective functions based on crystallographic statistics and generating crystal structures under symmetry constraints without relying on interatomic interaction models. MAGUS \cite{han2025efficient} accelerates evolutionary crystal structure prediction by explicitly exploiting group-subgroup relations and symmetry-kept mutation operators. Target XXXI \cite{zhou2025robust} focuses on robust polymorph discovery for small-molecule crystals by combining symmetry-aware packing searches with the use of machine learning
force fields in a hierarchical crystal energy ranking. MACS \cite{zamaraeva2025macs} formulates periodic crystal geometry optimization as a multi-agent reinforcement learning problem, treating atoms as cooperative agents that iteratively minimize atomic forces using decentralized policies guided by local geometric observations and force-based rewards.}

\textcolor{black}{Overall, as discussed in Sections 4.1 and 4.2,
property prediction models and crystal structure generative models exhibit substantial overlap in their technical foundations. Both typically adopt similar data representations, such as graph-based representations of atomic structures, and rely on comparable backbone architectures, such as GGNNs. Owing to this shared foundation, property-prediction models are often integrated into generative frameworks as evaluators to assess the quality and functionality of generated structures.
Despite these similarities, the two paradigms differ fundamentally in their objectives and learning formulations. Property prediction models are designed to learn a forward mapping from a given crystal structure to target physical or chemical properties, with an emphasis on predictive accuracy and generalization. In contrast, crystal structure generative models aim to learn the underlying distribution of crystalline structures themselves, enabling the discovery of novel, diverse, and potentially stable materials that may not be present in existing datasets.
}

\subsection{Aiding Characterization}

Aiding characterization techniques, including XRD, SEM, SPM, and TEM, aim to provide insights into the microstructure, composition, and defects within materials, thereby facilitating the correlation between microstructural characteristics and their properties. In this part, we outline image-based aiding characterization and spectra-based aiding characterization. A more comprehensive review regarding the two aiding characterization techniques can be found in REF.\cite{choudhary2022recent}.


The atomistic images obtained by microscopy techniques can unveil their structural characteristics from microstructure to macrostructure. These structural characteristics are intimately connected to the materials' functionality and performance. Therefore, with the atomistic images, a broad portfolio of image-based material characterization techniques has been proposed to identify structural characteristics. In what follows, we briefly introduce some representative works in this field.

Ziletti \textit{et al.}~\cite{ziletti2018insightful} paved the way for this application by creating a large database of perfect crystal structures. In particular, by training on the database, the proposed ConvNet uses CNNs to classify the space group of each diffraction pattern. 
Image-based characterization techniques have also been used to generate predictions for every pixel in an image, providing detailed information about the size, position, orientation, and morphology of features. Azimi \textit{et al.}~\cite{azimi2018advanced} developed an ensemble of fully CNNs to accurately segment martensite, tempered martensite, bainite, and pearlite in SEM images of carbon steels, achieving a remarkable accuracy of 94\%. The methodology represents an advancement in automating the segmentation of different phases in SEM images. Yang \textit{et al.}~\cite{yang2021deep} utilized the U-net architecture to attain high accuracy in detecting vacancies and dopants in scanning transmission electron microscopy images, achieving a model accuracy of up to 98\%. To classify these atomic sites, they grounded their approach in experimental observations, categorizing the sites into five distinct types: tungsten, vanadium replacing tungsten, selenium with no vacancy, monovacancy of selenium, and divacancy of selenium.

When electromagnetic radiation hits materials, the interaction between the radiation and the substance can be measured by the wavelength or frequency of the radiation --- spectra signals. The spectra data are commonly obtained by XRD spectra and Raman spectra~\cite{el2019raman, lee2020deep,li2024database,choudhary2022recent}. XRD spectra, represented as plots of X-ray intensity versus the angle between the incident and diffracted beams, provide quantitative information about crystal structures, such as the positions and intensities of diffraction peaks, which correspond to interplanar spacings and crystallite sizes. Raman spectra, obtained by measuring the frequency shift of scattered light, provide characteristic information about molecular vibrations and crystal structures. Raman spectra can identify the chemical composition and crystal structure of materials.

Park \textit{et al.}~\cite{park2017classification} analyzed 150,000 XRD patterns and utilized CNN models to predict structural information from the simulated patterns. The CNN models achieved accuracies of 81.14\%, 83.83\%, and 94.99\% for space-group, extinction-group, and crystal-system classifications, respectively. Abhiroop \textit{et al.}~\cite{bhattacharya2022deep} proposed a framework using CNNs and long short-term memory networks for the automatic identification of complex polymorph structures from their Raman spectra. This framework can accurately recognize TiO2 with different polymorphs. Stein \textit{et al.}~\cite{stein2019synthesis} explored the mapping between material images and their corresponding spectra using a conditional variational encoder, employing neural network models as the backbone. The model can generate spectra directly from straightforward material images, enabling significantly faster material characterizations.

\subsection{Accelerating Theoretical Computations}


Theoretical computations serve as indispensable tools for understanding the behavior of molecules and materials at the atomic and molecular levels.
In this section, we present the recently proposed machine learning-assisted force field development and quantum mechanics for accelerating theoretical computations.


\subsubsection{Force Field Development}


Force fields or interatomic potentials are empirical models that use potential energy functions and parameters to describe the atomic interactions and dynamics for atomistic simulations. By learning accurate force fields, researchers can precisely model the behavior of materials at the atomic scale, thus understanding the property-structure relationships. Empirical force fields often rely on hand-crafted parameters, which are efficient yet inaccurate. On the other hand, the estimation of energy functions and parameters can be obtained by extensive optimization to fit the data from quantum mechanical calculations. They are accurate but inefficient. To strike a balance between accuracy and efficiency, machine learning-based force field methods have been proposed to model the force fields from calculation and experiment data.

The pioneer works of machine learning-based interatomic potentials leveraged descriptor-based frameworks with shallow neural networks or Gaussian Processes~\cite{behler2007generalized,bartok2010gaussian,batzner20223}. Behler \textit{et al.}~\cite{behler2007generalized} proposed the first neural network potential, BPNN. In BPNN, an atomic descriptor (i.e., basis functions that transform an atomic configuration into a fixed-length fingerprint vector) based on the bond lengths and bond angles is passed to an MLP. On top of this formulation, a Monte Carlo dropout technique can be applied to the MLP to equip the potential with the ability to quantify its predictive uncertainty. Based on BPNN, various variants have been developed for molecular systems, such as water, methane, and other organic molecules~\cite{behler2007generalized,ko2021fourth}.

Recently, GGNNs have emerged as powerful deep-learning techniques for interatomic potentials due to their superior ability to incorporate invariant/equivariant symmetry constraints and long-range interaction~\cite{deng2023chgnet}. They eliminate the need for hand-crafted descriptors. Cormorant~\cite{anderson2019cormorant} uses an equivariant neural network for property prediction on small molecules. This method is demonstrated on the potential energies of small molecules but not on atomic forces or systems with periodic boundary conditions. PotNet~\cite{lin2023efficient} models interatomic potentials by including the Coulomb potential, London dispersion potential, and Pauli repulsion potential. PotNet employs a message-passing scheme that considers interatomic potentials as well as efficient approximations to capture the complete set of potentials among all atoms. PotNet outperforms other methods on all four tasks, including the prediction of formation energy, band gap, bulk moduli, and shear moduli with the mean absolute error of 0.0188, 0.204, 0.040, and 0.065, respectively, on the Materials Project dataset.

Most previous GGNN-based force fields are based on two-atom interactions, known as two-body information, which means they only rely on the states of two atoms, known as two-body interactions. This is not in line with the actual situation, i.e., multiple atoms interact with each other. To address this limitation, MACE~\cite{batatia2022mace}, an architecture combining equivariant message passing with the efficient many-body message, aims to capture four-body interactions. MACE shows that using four-body messages reduces the required number of message passing iterations to just two, resulting in a fast prediction, reaching state-of-the-art accuracy of energy and force prediction on the small molecule datasets.
In addition, DimeNet~\cite{gasteigerdirectional} enhances the use of pairwise interactions in a single convolution by integrating angular and three-body terms. DimeNet achieves a mean absolute error of 0.035 kcal mol$^{-1}$ for energy, 0.07 kcal mol$^{-1}$ for energy, and 0.17 kcal mol$^{-1}$ \AA$^{-1}$ for forces on the MD17 dataset~\cite{doiStefan1603015}, surpassing the performance of alternative GGNN models.

Towards the generalizability across the periodic table, Chen \textit{et al.} used more than 187,000 energies, 16,000,000 forces, and 1,600,000 stresses from relaxation structures to train a universal deep learning framework for interatomic potentials, called M3GNet~\cite{chen2022universal}. As shown in FIG. \ref{fig:overview-generation}e. Firstly, many-body computation calculates the bond angles $\theta$ and $\tau$ between three atoms and many atoms, respectively. Then, the many-body to bond module leverages the angles $\theta$, $\tau$, and bond distance to construct bond information $e^{'}$. Afterward, a standard graph convolution updates bond and atom information by using bond information $e^{'}$ and atom information $v$. This many-body interaction considers richer angle information for material modeling, improving the performance of force field modeling. As shown in FIG. \ref{fig:overview-generation}f and FIG. \ref{fig:overview-generation}g, the energy predicted by M3GNet is very close to the energy calculated by DFT. M3GNet showcases the potential to learn the interatomic potential of various materials across the periodic table. As shown in FIG. \ref{fig:overview-generation}h, for any chemistry (All) and oxides only (Oxides), the predicted signed $E_{hull}$ distribution by M3GNet is close to DFT calculation. The sable ratio of the top-1000 lowest $E_{hull-m}$ materials from All and Oxides is reported in FIG. \ref{fig:overview-generation}i.


Although machine learning-based force field development has achieved the balance between efficiency and accuracy, they heavily rely on the large-scale labeled training data from quantum mechanics calculations, e.g., DFT. This computationally expensive data collection process has become a bottleneck in developing force fields. To reduce the cost of DFT labeled samples, Shui \textit{et al.}~\cite{shui2022injecting} proposed two strategies to achieve weakly supervised learning of neural network potentials by utilizing physical information in empirical force fields. The first strategy is to train a classifier to select the best possible empirical force field for unlabeled samples and use the energy value calculated from this force field as the label value for the unlabeled sample to achieve data augmentation. The second strategy is based on transfer learning, which first trains on a large dataset obtained through empirical force fields and then uses DFT-labeled samples for fine-tuning. The experiment shows that the first strategy can improve performance by 5\%$\sim$51\%, while the second strategy can improve performance by up to 55\%. 

With the same goal of improving the data efficiency, NequIP~\cite{batzner20223} uses relative position vectors instead of simple distances (scalars), which not only contain scalars but also features of high-order geometric tensors. This changes the features to be rotation invariant and allows rotation invariant filters to use angle information. It employs E(3)-equivariant convolutions for interactions of geometric tensors, resulting in state-of-the-art accuracy and data efficiency. NequIP is applied to a catalytic surface reaction of heterogeneous catalysis of formate dehydrogenation, and it can obtain mean absolute errors in the force components of 19.9 meV/\AA, 71.3 meV/\AA, 13.0 meV/\AA, and 47.6 meV/\AA, on the four elements C, O, H, and Cu, respectively, as well as an energy mean absolute error of 0.50 meV/atom, demonstrating that NequIP is able to accurately model the interatomic forces for this complex reactive system.

On the other hand, the ion’s valence can engage in different bonding depending on the number of electrons and the environment. Therefore, incorporating the impact of valence on chemical bonds into neural network potentials is very important. Traditional neural network potentials treat the elemental label as the basic chemical identity, but different valence states of transition-metal ions behave differently from each other as different elements. To narrow the research gap, CHGNet~\cite{deng2023chgnet} takes a crystal structure with unknown atomic charges as input and outputs the corresponding energy, forces, stress, and magnetic moments (magmoms). The charge-decorated structure can be inferred from the on-site magmoms and atomic orbital theory. CHGNet regularizes the node-wise features at the convolution layer to contain the information about magmoms. The regularized features carry rich information about both local ionic environments and charge distribution. Therefore, the atom features used to predict energy, force, and stress are charge-constrained by their charge-state information. As a result, CHGNet can provide charge-state information using only the nuclear positions and atomic identities as input, allowing the study of charge distribution in atomistic modeling. 
CHGNet is pre-trained using energies, forces, stresses, and magnetic moments obtained from the Materials Project Trajectory Dataset~\cite{deng2023chgnet} to learn the orbital occupancy of electrons, thereby enhancing its ability to describe both atomic and electronic degrees of freedom. Its effectiveness has been demonstrated in solid-state materials, including charge-informed molecular dynamics in LixMnO2, the finite temperature phase diagram for LixFePO4, and lithium diffusion in garnet conductors.


Despite the progress, the current evaluation criteria for machine learning force fields are limited to the prediction accuracy of force and energy. Fu \textit{et al.}~\cite{fu2023forces} selected a series of systems, including water, organic small molecules, peptides, and crystal materials, and designed a series of evaluation criteria to describe trajectory stability, demonstrating that some machine learning force fields with high accuracy currently cannot reproduce dynamics trajectories well. The work suggested that stability should become a new criterion for evaluating machine learning force fields.

\subsubsection{Aiding Quantum Mechanics}

Force field development heavily relies on empirical data and may struggle with accuracy in complex systems. On the other hand, rigorous quantum mechanical approaches, e.g., DFT, provide a more accurate description of electronic structure within molecules and materials systems and thus accurately predict their behavior and properties. While DFT excels in accuracy, it is computationally demanding, especially for large systems, which can limit its practical applications.

To improve the efficiency of DFT, the tight-binding model provides a more streamlined approach. Tight-binding model is based on the assumption that electrons are tightly bound to their respective atoms and only interact with their nearest neighbors~\cite{wang2021machine,agapito2016accurate,smutna2020systematic}. By using this approximation, the tight-binding model significantly reduces the computational cost while maintaining a level of accuracy. Building upon tight-binding methods, Density Functional Tight Binding (DFTB) emerges as an advanced method that combines the advantages of DFT with the efficiency of tight binding~\cite{li2022deep}. DFTB utilizes parameterized interactions derived from DFT calculations, allowing it to capture essential chemical properties while accelerating the computational process. This makes DFTB particularly useful for studying larger systems, such as complex materials and biomolecular structures~\cite{gu2024deep,stöhr2020accurate,wang2021machine}.

Recently, incorporating machine learning into DFTB has further advanced the field~\cite{wang2021machine,gu2022neural,gu2024deep}. They mostly learn from a vast dataset obtained by quantum mechanical calculations to learn the mapping from structure to Hamiltonians which is a bedrock property in DFT calculations.
Wang~\textit{et al.} developed machine learning-based algorithms to generate tight-binding Hamiltonians matrices from electronic eigenvalues~\cite{wang2021machine}. However, this approach neglected atomic structure information, limiting its applicability to "unseen" structures. 
Recently, a deep learning-based tight-binding method, dubbed DeePTB, is proposed~\cite{gu2024deep}. DeePTB constructs tight-binding Hamiltonians using gauge-invariant parameters and then maps these parameters from symmetry-preserving local environment descriptors to obtain the tight-binding Hamiltonian and its corresponding eigenvalues. After supervised learning from training structures with ab initio eigenvalues, DeePTB can directly predict accurate tight-binding Hamiltonians. On the group IV and group III-V materials, exceptional agreement of eigenvalues from DeePTB Hamiltonians and those from ab initio calculations can be achieved, with the coefficient of determination $R^2\approx 0.9999$.

Despite the progress, it should be pointed out that these methods are still mostly based on data-driven inference instead of physical laws. The physical laws governed by quantum mechanics indeed determine the atomic interactions. To address the issue, Stohr \textit{et al.} proposed a hybrid quantum mechanics/deep learning formalism, called the DFTB-NN$_{rep}$~\cite{stöhr2020accurate}. DFTB-NN$_{rep}$ uses DFTB in conjunction with a global deep neural network. Specifically, the repulsive energies and forces are obtained via a neural network model based on reference data, while the electronic energies and properties are calculated by DFTB.
DFTB-NN$_{rep}$ has demonstrated highly accurate predictions of energetic, structural, and vibrational properties for a vast range of organic molecules across their respective conformational spaces.

\section{Benchmarking and Software Platforms}
The availability of volume datasets~\cite{jain2013commentary,choudhary2020joint,saal2013materials,kirklin2015open,castelli2012new,carbon2020data,vaitkus2021validation,el2019raman,bergerhoff1987crystallographic,zagorac2019recent,zitnick2020introduction,petretto2018high,zhao2021high,deng2023chgnet,schmidt2021crystal,rosen2021machine,barroso2024open,chenlearning,Atomly,LeMaterialDatasets2025}, \textcolor{black}{benchmarks}~\cite{dunn2020benchmarking,du2023m,choudhary2024jarvis,omee2024structure,bin2025simxrd,chenecd,alampara2025probing,riebesell2025framework}, and platforms is the fuel in the field of AI-driven materials science. In this section, we introduce commonly utilized datasets, \textcolor{black}{benchmarks}, and software platforms~\cite{ong2013python,choudhary2020joint} in crystalline material research. \textcolor{black}{Due to page limits, the related content is provided in the Supplemental Material.}

\section{Conclusions and Future Outlooks}

In conclusion, the integration of AI, particularly deep learning, has transformed crystalline material discovery. By leveraging advanced data representations and models such as geometric graph neural networks, transformers, and diffusion models, researchers can predict material properties, synthesize new materials, aid characterization, and accelerate theoretical computations, with exceptional efficiency and accuracy. However, significant challenges remain. In the following section, we will outline future directions to further unlock the potential of AI in this field.

\subsection{Self-Driving Laboratories}

In materials science, traditional experimental methods often involve researchers manually designing and conducting experiments, which can be time-consuming and labor-intensive~\cite{cawse2003experimental,talapatra2018autonomous}. In contrast, the recently proposed self-driving or autonomous laboratories use machine learning models to plan and perform experiments, increasing laboratory automation and accelerating the scientific discovery processes~\cite{burger2020mobile,pyzer2022accelerating}.In this section, we first highlighted representative works on self-driving laboratories and provided further clues.

In these years, self-driving laboratories have covered the automation of retrosynthesis analysis (such as in reinforcement-learning-aided synthesis planning~\cite{coley2019robotic}), prediction of reaction products (such as in CNNs for reaction prediction~\cite{coley2019graph}) and reaction condition optimization (such as in robotic workflows optimized by active learning~\cite{burger2020mobile}). Recently, LLMs have provided valuable insights to autonomous driving laboratories, thanks to their extensive knowledge of physics and chemistry, as well as their abilities in chain-of-thought reasoning, self-reflection, and decision-making.~\cite{ramos2024review}. These LLMs as autonomous agents are typically augmented with tools or action modules, empowering them to directly interact with the physical world, e.g., scientific experimentation~\cite{m2024augmenting,jia2024llmatdesign}. 

One representative work is ChemCrow~\cite{m2024augmenting}. By combining the reasoning power of LLMs with chemical expert knowledge from 18 computational tools in chemistry, ChemCrow showcases one of the first chemistry-related LLM agent interactions with the physical world. ChemCrow has planned and synthesized an insect repellent and three organocatalysts and guided the screening and synthesis of a chromophore with target properties. Jia \textit{et al.} introduced LLMatDesign, a LLM-based framework for materials design~\cite{jia2024llmatdesign}. LLMatDesign utilizes LLM agents to translate human instructions, apply modifications to materials, and evaluate outcomes using provided tools. By incorporating self-reflection on its previous decisions, LLMatDesign adapts rapidly to new tasks and conditions in a zero-shot manner. An evaluation of LLMatDesign on several materials design tasks in silico validates LLMatDesign's effectiveness in developing new materials with user-defined target properties in the small data regime. The pioneering work mentioned above has established the foundational paradigms for self-driving laboratories. However, numerous opportunities remain for further research and development.

1) Human–AI collaboration: First, to prevent self-driving experiments from involving unethical purposes or high-risk operations, humans must validate AI agents' operations, ensuring they align with accepted scientific principles, safety, and ethical standards~\cite{ji2024beavertails,yao2024survey}. Second, experiments often lead to unexpected results that require immediate rethinking and adjustments to protocols. In this scenario, human judgment is invaluable~\cite{liu2023conversational,drori2024human,huang2024replacing}. By involving human experts in the research loop, their knowledge can be leveraged to inform the design and refinement of experiments, and the interpretation of findings~\cite{liu2022machine,tiihonen2022more}. Therefore, LLMs could play an indispensable role in this "humans in the loop" process, as they provide conversational interfaces that allow researchers to incorporate feedback and human experience.


2) Configurable module integration: Self-driving laboratories should automatically select suitable tools from a wide range of computing and experimental tools and configure them to perform specific tasks~\cite{abolhasani2023rise,kannan2023can}. Also, the optimal parameters and configurations of these tools should be automatically determined. In addition, for certain experimental processes, such as material screening, many algorithms exist for executing this process. Therefore, it is necessary to conduct algorithm selection to select the most suitable one to perform~\cite{kerschke2019automated}. These challenges may provide an opportunity for the future development of self-driving laboratories toward standardized operation modes and end-to-end experimental workflows.

3) Material language processing: Knowledge in materials science is often recorded in the form of text language, including not only structured data such as CIFs and SLICES but also non-structured data like literature, descriptions, etc. Therefore, effectively processing and learning these different types of material languages can further unleash the potential of AI in material discovery automation.

\subsection{Generalizability}

Deep learning could suffer from generalizability issues in material science, i.e., their performance on new tasks or unseen data will degrade. To improve the generalizability, one straightforward way is to pre-train on large-scale databases. For example, M3GNet~\cite{chen2022universal} utilizes the largely untapped dataset of more than 187,000 energies, 16,000,000 forces, and 1,600,000 stresses from structural relaxations performed by the Materials Project to train the graph deep learning model. M3GNet showcases the potential generalizability to predict the formation energy, etc, of various materials across the periodic table.

However, pre-training techniques require a large amount of data and computational resources. Furthermore, the distribution of new data may shift from the pre-training data distribution, i.e., out-of-distribution~\cite{omee2024structure}, which could degrade the performance under the new data. In addition to the out-of-distribution caused by new samples, this bias could also stem from the following two aspects: First, crystalline materials can exhibit different behaviors under extreme conditions such as high temperature, pressure, or strain~\cite{wang2024comprehensive}. Models trained on data within normal ranges might fail to generalize to these extreme conditions. Second, crystal defects, such as vacancies, interstitials, or dislocations, introduce variations that might not be adequately captured by models trained on defect-free structures, contributing to the generalizability challenge~\cite{ziletti2018insightful}. To improve the generalizability under out-of-distribution samples, more advanced pretraining techniques~\cite{wang2024comprehensive},
transfer learning~
\cite{9336290}, retrieval augmented generation~\cite{gao2023retrieval}, and zero-shot learning techniques~\cite{10582688shotLearning} can be developed in the future.

\subsection{\textcolor{black}{Evaluation Metrics for Ab Initio Crystal Generation}}

\textcolor{black}{
For \emph{ab initio} crystal generation tasks, a high-fidelity evaluation of the stability of generated materials ideally requires quantum-mechanical (QM) calculations. However, using QM methods directly for large-scale evaluation is computationally prohibitive. As a result, early works such as CDVAE \cite{xie2021crystal} primarily relied on low-cost but physically meaningful metrics, including \emph{Validity}, \emph{Coverage}, and \emph{Property statistics}, to assess the performance of crystal generative models. These metrics mainly measure the similarity between generated samples and the test set, or verify basic structural validity.
Nevertheless, the practical discovery of new materials requires not only valid structures but also stable and diverse materials. To address this limitation, MatterGen \cite{zeni2025generative} introduced the \emph{Stability, Uniqueness, and Novelty (S.U.N.)} metrics for more comprehensive evaluation, which combine DFT calculations with machine-learned interatomic force fields. Following this trend, future research is likely to develop additional low-cost evaluation metrics tailored to specific objectives. For example, S.U.N. explicitly targets stability and diversity, while FlowMM further proposes the \emph{S.U.N. Cost} metric to quantify the expense of discovering new stable materials. WyFormer introduces symmetry-related metrics such as the proportion of P1 structures and the space group $\chi^2$ statistic to assess the model. In addition, SymmCD \cite{levysymmcd} and SGEquiDiff \cite{chang2025space} evaluate the fraction of unique and novel templates (U.N. templates), where a template is defined as a space group and multiset of occupied Wyckoff positions.
It is worth noting that none of these metrics can fully and accurately evaluate material performance. Even QM calculations themselves may fail to capture certain properties precisely \cite{zeni2025generative}. Ultimately, more rigorous validation often requires real experimental verification, which incurs significantly higher costs. In practice, computational methods and evaluation metrics are therefore commonly used to screen large numbers of generated candidates, after which only a small subset of promising samples is selected for experimental validation.
}

\subsection{\textcolor{black}{Energy Underestimation in Machine-Learning Potentials}}

\textcolor{black}{Recent studies have identified a systematic energy underestimation phenomenon in modern machine-learning interatomic potentials (MLIPs), particularly for out-of-distribution and high-energy configurations, which manifests as a softening of the learned potential energy surface (PES)~\cite{deng2025systematic}. From a machine learning perspective, this issue is closely related to the long-standing problem of class and distribution imbalance, as most training datasets are dominated by near-equilibrium crystal structures, while strained states, defects, surfaces, and transition configurations are severely underrepresented. As a result, MLIPs tend to achieve low average errors on in-distribution samples while exhibiting biased predictions in sparsely sampled, high-energy regions of configuration space. This imbalance also complicates evaluation, since standard benchmarks and random test splits often fail to expose systematic PES softening. 
Therefore, future research directions include constructing datasets with improved coverage of high energy regions, designing benchmarking protocols that explicitly probe distributional imbalance, and developing data efficient mitigation strategies such as active learning to correct systematic energy bias. These efforts are essential for improving the robustness and reliability of MLIPs in crystal modeling beyond equilibrium structures.
}

\subsection{\textcolor{black}{AI for Disordered Crystalline Materials}}
\textcolor{black}{
Most existing AI models for crystalline materials primarily focus on ideal ordered periodic structures. In contrast, a substantial fraction of experimentally realized materials exhibit various forms of crystallographic disorder, including substitutional disorder and positional disorder \cite{petersen2025dis,chenlearning,jakob2025learning}. Disordered crystals are particularly important in practical applications, as disorder is ubiquitous in functional materials where ion mixing, alloying, and doping strategies are commonly employed to enhance stability or achieve targeted properties \cite{petersen2025dis}.
However, current AI approaches devote limited attention to disordered crystals. Only a small number of models explicitly consider disordered crystal structures. In most cases, disorder is treated as a static structural attribute, for example, by using averaged occupancies or a single representative configuration, without explicitly modeling the underlying configurational ensembles in a principled manner.
Developing more faithful representations and learning frameworks for disordered crystals, enabling accurate property prediction and generative modeling, and constructing MLIPs for disordered systems remain critical open challenges. Addressing these issues is essential for narrowing the gap between idealized AI based crystal modeling and experimentally realizable materials.
}

\subsection{Explainability}

To realize the full potential of AI for scientific discovery, AI should not only have prediction and generation abilities but also need explainability. This explainability fosters trust in model outputs, facilitates the validation of scientific hypotheses, and offers actionable insights for refining research investigations, such as optimization for molecular and material structural design~\cite{wang2024explainable,gallegos2024explainable}. Although many fundamental works have explored interpretability in materials science~\cite{wang2024comprehensive,vinchurkar2024explainable,xie2018crystal,wang2024explainable} many challenges remain unresolved. 

1) Explainability granularity: Due to complex and non-linear relationships between material properties and features, material properties can depend on features at different granularity, including atomic, microstructural, and macroscopic structures. However, research on obtaining explanations at various granularities is still very limited.

2) Explainability robustness: Although there are numerous interpretable methods available, such as attention-based, gradient-based, perturbation-based, etc., they may not necessarily produce robust and faithful explanations~\cite{agarwal2021towards,wang2024explainable}. In addition, experimental data often contains noise or discrepancy from measurement uncertainties and environmental conditions. In these scenarios, unfaithful explanations could be inferred and result in misleading scientific clues. 

3) Interdisciplinary knowledge requirement: Understanding materials science requires knowledge of physics and chemistry, making it challenging to create interpretable models that are accessible to all researchers from different fields. Therefore, relevant benchmarks and ground truth are needed to promote the development of interpretable models in this interdisciplinary.

\newpage

\section*{Supplemental Materials}


\subsection*{Geometric Graph Neural Networks}

GGNNs are specifically designed to handle the complex geometric relationships within the geometric graph data, such as proteins, molecules, and crystalline materials~\cite{han2024survey,xie2021crystal}. The key design principles of GGNNs are rooted in the principles of message-passing and aggregation mechanisms. 
The message-passing mechanism processes graph-structured data by exchanging each node's information with its neighboring nodes through edges. The aggregation performs local updates to node features by aggregating information from its immediate neighbors.
Formally, the entire process can be expressed as follows \cite{gilmer2017neural}:
\begin{equation}\label{eq:egnn}
m_{ij}=\phi_{\mathrm{msg}}\left(h_{i}^{(l-1)},h_{j}^{(l-1)},d_{ij},e_{ij}\right),\quad
h_{i}^{(l)}=\phi_{\mathrm{agg}}\left(h_{i}^{(l-1)},\{m_{ij}\}_{j\in\mathcal{N}_{i}}\right),
\end{equation}
where $\phi_{\mathrm{msg}}(\cdot)$ and $\phi_{\mathrm{agg}}(\cdot)$ represent the message calculation and aggregation, respectively, $h_{i}^{(l-1)}$ and $h_{j}^{(l-1)}$ are the node features of the $i$-th and $j$-th nodes in the $(l-1)$-th layer, respectively, $h_{i}^{(l)}$ represents the node feature of the $i$-th node in the $l$-th layer, $d_{ij}=||x_i-x_j||^2$ is the directional information between the node coordinates $x_i$ and $x_j$, $e_{ij}$ denotes the edge features, and $m_{ij}$ represents the computed message. 

Noticeably, compared to the vanilla message-passing mechanism in graph neural networks, Eq. (\ref{eq:egnn}) also encodes the directional information $d_{ij}$ between two nodes (angle information can also be included). By integrating this geometric information, GGNNs can perceive the symmetry inherent in crystalline materials. This capability allows GGNNs to preserve equivariance or invariance to group transformations such as translations, rotations, and reflections, which ultimately enhances the model's generalization ability and robustness.
Box 1 provides the concepts of group transformations, equivariance, and invariance, respectively.

\begin{tcolorbox} [breakable,boxrule=.2mm,arc = 0mm, outer arc = 0mm,left=0.2mm,right=0.2mm,top=0.3mm,bottom=0.3mm]
\noindent \textbf{Box 1 | Group, Equivariance and Invariance.}\tcblower

\textbf{Group}

A group $G$ is a set of transformations with a binary operation "$\cdot$" that satisfies the following properties: 

(i) Closure: $\forall a,b\in G,a\cdot b\in G$; 

(ii) Associativity: $\forall a,b,c\in G,(a\cdot b)\cdot c=a\cdot(b\cdot c)$; 

(iii) Identity element: there exists an identity element $e\in G$ such that $\forall a\in G,a\cdot e=e\cdot a=a$; 

(iv) Inverses: there exists an identity element $e\in G$ such that $\begin{aligned}\forall a\in G,\exists b\in G,a\cdot b=b\cdot a=e\end{aligned}$, where the inverse $b$ is denoted as $a^{-1}$. 

Here are some groups commonly used in crystalline materials research.
\begin{enumerate}
\item $E(d)$ is an Euclidean group comprising rotations, reflections, and translations, acting on $d$-dimension vectors.
\item $O(d)$ is an orthogonal group,  consisting of rotations and reflections, acting on d-dimension vectors.
\item $SE(d)$ is a special Euclidean group, which includes only rotations and translations.
\item $SO(d)$ is a special orthogonal group, consisting exclusively of rotations.
\item $S_N$ is a permutation group, whose elements are permutations of a given set consisting of $N$ elements.
\end{enumerate}

\textbf{Equivariance}\\ A function $f: \mathcal{X} \to \mathcal{Y}$ is $G$-equivariant if applying a group transformation to the input results in the same transformation applied to the output:
$\phi(g \cdot x) = g \cdot \phi(x), \forall g \in G$. With group representation, this becomes:
$\phi(\rho_{\mathcal{X}}(g)x) = \rho_{\mathcal{Y}}(g)\phi(x), \forall g \in G$, where $\rho_{\mathcal{X}}(\cdot)$ and $\rho_{\mathcal{Y}}(\cdot)$ are the group representations in the input and output spaces, respectively. 

\textbf{Unit Cell E(3) Equivariance}\\ For crystalline materials, if $f: (\mathbf{A}, \mathbf{X}, \mathbf{L}) \to \mathcal{Y}$ is unit cell $E(3)$-equivariant, then:
$\mathbf{Q}f(\mathbf{A}, \mathbf{X}, \mathbf{L}) = f(\mathbf{A}, \mathbf{Q} \mathbf{X} + \boldsymbol{b}, \mathbf{Q} \mathbf{L})$, where $\mathbf{Q}$ is a rotation matrix and $\boldsymbol{b}$ is a translation vector.

\textbf{Permutation Equivariance}\\ 
If function $f:(\mathbf{A},\mathbf{X},\mathbf{L})\to\mathcal{Y}\in \mathbb{R}^n$ is permutation equivariant, then we have $\mathbf{P}f(\mathbf{A},\mathbf{X},\mathbf{L}) = f(\mathbf{P}\mathbf{A},\mathbf{P}\mathbf{X},\mathbf{L})$, where $\mathbf{P}\in\{0,1\}^{n\times n}$ is a permutation matrix (represent the operation of adjusting the atom order). 

\textbf{Invariance}\\ Similar to equivariance, a function $f: \mathcal{X} \to \mathcal{Y}$ is $G$-invariant if applying a group transformation to the input leaves the output unchanged:
$\phi(g \cdot x) = \phi(x), \forall g \in G$. With group representation, this becomes:
$\phi(\rho_{\mathcal{X}}(g)x) = \phi(x), \forall g \in G$.

\textbf{Unit Cell E(3) Invariance}\\ We particularly consider the invariance of unit cells. For a function $f:(\mathbf{A},\mathbf{X},\mathbf{L})\to\mathcal{Y}$, if it is invariant under the $E(3)$ group, then we have $f(\mathbf{A},\mathbf{X},\mathbf{L}) = f(\mathbf{A},\mathbf{Q}\mathbf{X}+\boldsymbol{b},\mathbf{Q}\mathbf{L})$, where $\mathbf{Q}\in\mathbb{R}^{3 \times 3}$ is a rotation and reflection matrix and $\boldsymbol{b}\in\mathbb{R}^{3}$ is a translation vector. 

\textbf{Permutation Invariance}\\ Similarly, we consider the permutation invariance w.r.t the atom order. 
If function $f:(\mathbf{A},\mathbf{X},\mathbf{L})\to\mathcal{X}$ is permutation invariant, then we have $f(\mathbf{A},\mathbf{X},\mathbf{L}) = f(\mathbf{P}\mathbf{A},\mathbf{P}\mathbf{X},\mathbf{L})$, where $\mathbf{P}$ is a permutation matrix. 

\textbf{Periodic Invariance}\\ \textcolor{black}{For periodic crystal data, different choices of unit cell representation, such as integer supercell constructions or shifts of the unit cell origin, can lead to different representations of the same infinite crystal.}
Consequently, models applied to crystal data also need to be periodic invariant. If a function $f:(\mathbf{A},\mathbf{X},\mathbf{L})\to\mathcal{X}$ exhibits periodic invariance,  $f(\mathbf{A},\mathbf{X},\mathbf{L})=f(\Phi(\hat{\mathbf{A}},\hat{\mathbf{X}},\boldsymbol{\alpha}\mathbf{L},\mathrm{p}),\boldsymbol{\alpha}\mathbf{L})$, where $\Phi:(\hat{\mathbf{A}},\hat{\mathbf{X}},\mathbf{L},\mathrm{p})\to(\mathbf{A},\mathbf{X})$ represents the abstract generating function of the unit cell and $\mathrm{p}$ is a corner point. $\alpha\in\mathbb{N}_{+}^{3}$ is a scalar that pertains to the scaling of unit cell size.

\end{tcolorbox}

\subsection*{Convolutional Neural Networks}

CNNs have revolutionized the field of deep learning, particularly in vision domains. In recent years, CNNs have also been applied to atomic image-based aiding characterization tasks such as structure recognition, defect detection, and microstructure analysis~\cite{ziletti2018insightful,azimi2018advanced,yang2021deep}. A typical CNN architecture consists of several key components, including convolution layers, pooling layers, activation functions, and fully connected layers.

Convolutional layers aim to autonomously detect spatial hierarchies within data. Specifically, convolutional operations involve sliding a convolutional kernel over the input data to compute local patterns. This process enables the CNNs to learn features from small local regions and progressively capture more complex patterns as the network deepens. Following the convolutional layers, pooling layers are used to downsample the feature maps, retaining the most critical information while reducing dimensionality. Common pooling methods include max pooling and average pooling, both of which help mitigate overfitting and decrease computational load. CNNs incorporate activation functions like ReLU to introduce non-linearity into the model. In standard CNN architectures, the final layers are fully connected layers, which use the high-level features learned by the convolutional and pooling layers to make final predictions.

\subsection*{Language Models}

The evolution of language models within the domain of deep learning has witnessed remarkable progress over the past decade, particularly in the field of natural language processing~\cite{han2022survey}.
With the emergence of chemical languages, e.g., simplified molecular input line entry systems (SMILES~\cite{weininger1988smiles}), language models have also demonstrated tremendous potential in the field of science. 
Current mainstream language models primarily rely on transformer architectures which consist of an encoder and a decoder. 
The encoder's role is to convert an input sequence into a fixed-length vector, or embedding. This embedding serves as a comprehensive representation, encapsulating both short- and long-range dependencies within the data. The decoder subsequently leverages this embedding to generate the output sequence.

The core innovation of the transformer is the self-attention mechanism. This allows the model to weigh the significance of different input tokens in a sequence relative to one another. By calculating attention scores, the model can focus on particular parts of the input, enabling it to capture contextual relationships without relying on sequential processing. Specifically, through defining three learnable weight matrices $\mathbf{W}_{Q}\in\mathbb{R}^{d\times d_q}$, $\mathbf{W}_{K}\in\mathbb{R}^{d\times d_k}$, and $\mathbf{W}_{V}\in\mathbb{R}^{d\times d_v}$,the input embeddings $\mathbf{X}\in\mathbb{R}^{n\times d}$ are linearly transformed to three parts, i.e., queries $\mathbf{Q}\in\mathbb{R}^{n\times d_q}$, keys $\mathbf{K}\in\mathbb{R}^{n\times d_k}$, values $\mathbf{V}\in\mathbb{R}^{n\times d_v}$, where $\mathbf{Q}=\mathbf{X}\mathbf{W}_{Q}$, $\mathbf{K}=\mathbf{X}\mathbf{W}_{K}$, $\mathbf{V}=\mathbf{X}\mathbf{W}_{V}$, and $d$, $d_q$, $d_k$, $d_v$ are the dimensions of inputs, queries, keys and values ($d_k=d_q$), respectively. The output of the self-attention layers is,
\begin{equation}
    \mathrm{attention}(\mathbf{Q},\mathbf{K},\mathbf{V})=\mathrm{softmax}\left(\frac{\mathbf{Q}\mathbf{K}^T}{\sqrt{d_q}}\right)\mathbf{V}.
    \end{equation}
Afterward, the output of the $H$ different self-attention layers is then concatenated and linearly transformed to multi-head attention output, $
\mathrm{MultiHeadAttnoutput}=\mathrm{Concat} (head_1,\cdots,head_H)\mathbf{W}_{Z}$,
where $head_i=\mathrm{attention}(\mathbf{Q_i},\mathbf{K_i},\mathbf{V_i})$, $\mathbf{W}_{Z}\in\mathbb{R}^{H\cdot d_v\times d}$.
Besides, a residual connection module followed by a layer normalization module is inserted around each module. That is, $\mathbf{H}^{\prime} = \mathrm{LayerNorm}(\mathrm{Attn}(\mathbf{X}) + \mathbf{X})$, $\mathbf{H} = \mathrm{LayerNorm}(\mathrm{FFN}(\mathbf{H}') + \mathbf{H}')$, where $\mathrm{Attn(\cdot)}$ denotes multi-head attention module,
$\mathrm{LayerNorm(\cdot)}$ denotes the layer normalization operation, and
$\mathrm{FFN(\cdot)}$ denotes the feed-forward network as
$\mathrm{FFN}(\mathbf{H}')=\mathrm{ReLU}(\mathbf{H}'\mathbf{W}_1+\mathbf{b}_1)\mathbf{W}_2+\mathbf{b}_2$, where $\mathbf{W}_1$, $\mathbf{W}_2$, $\mathbf{b}_1$, $\mathbf{b}_2$ are trainable parameters.

In recent years, the development of large language models (LLMs) based on the transformer has been driven by the availability of massive amounts of data and the increasing computational power. Based on the transformer architecture, LLMs, often with hundreds or even trillions of parameters, have achieved remarkable performance on a wide range of tasks, such as language processing and planning, scientific discovery\cite{wu2024evolutionary,zhang2024scientific}.


\subsection*{Diffusion Models}

Deep generative models have emerged as a transformative force in deep learning, enabling the generated data samples to resemble the real data~\cite{hong2024diffusion}. 
Traditional generative models such as Variational Autoencoders (VAE) and Generative Adversarial Networks (GANs) have significantly advanced data generation tasks. VAEs encode input data into latent variables and then decode the latent variables back to new samples. GANs, on the other hand, use an adversarial framework of two neural networks: a generator to synthesize data and a discriminator to distinguish whether samples are real or fake. 
However, these models often suffer from low-fidelity generative samples, mode collapse, or training instability~\cite{jabbar2021survey}.

Recently, diffusion models have exhibited advantages over traditional generative models by providing higher fidelity samples. 
Diffusion models generate data by gradually adding noise, namely the forward diffusion process, and reconstructing data by progressively removing the noise, namely the reverse diffusion process~\cite{yang2023diffusion,yang2023diffusion}. Specifically, the forward process incrementally adds noise to the real data $\mathbf{x}_0 \sim q(\mathbf{x}_0)$ over $T$ timesteps, 
\begin{equation}
   q(\mathbf{x}_{1:T}|\mathbf{x}_0)=\prod_{t=1}^{T} q(\mathbf{x}_{t}|\mathbf{x}_{t-1}), \quad
  q(\mathbf{x}_{t}|\mathbf{x}_{t-1})=\mathcal{N}(\mathbf{x}_{t};\sqrt{1-\beta_t}\mathbf{x}_{t-1},\beta_t\mathbf{I}).
\end{equation}
This phase is mathematically represented as a sequence of states, each representing a progressively noisier state compared to the preceding states. Conversely, the reverse diffusion process learns to reconstruct the original data $p_{\theta}(\mathbf{x}_0)$ by progressively removing the noise. 
\begin{equation}
   p_{\theta}(\mathbf{x}_{0:T})=p(\mathbf{x}_{T}) \prod_{t=1}^{T} p_{\theta}(\mathbf{x}_{t-1}|\mathbf{x}_{t}), \quad
   p_{\theta}(\mathbf{x}_{t-1}|\mathbf{x}_{t})=\mathcal{N}(\mathbf{x}_{t-1};\mu_{\theta}(\mathbf{x}_{t},t),\Sigma_{\theta}(\mathbf{x}_t,t)).
\end{equation}
The reverse diffusion process uses a neural network $\mu_{\theta}$ to recover the denoised state from the noisy input. Through the forward and reverse diffusion process, new data samples can be generated.

\subsection*{Benchmarking and Software Platforms}



The availability of substantial volume datasets, \textcolor{black}{benchmarks}, and platforms is the fuel in the field of AI-driven materials science. In this section, we introduce commonly utilized datasets, \textcolor{black}{benchmarks}, and platforms in crystalline material research.


\textcolor{black}{\textbf{Common Datasets}}
\begin{enumerate}

\item \href{https://next-gen.materialsproject.org/}{\textcolor{black}{\textit{The Materials Project} }}~\cite{jain2013commentary}:
The Materials Project (MP) harnesses the power of supercomputers alongside state-of-the-art quantum mechanics theories and DFT to systematically compute the properties of a vast array of materials. At present, the MP dataset encompasses over 120,000 materials, each accompanied by a comprehensive specification of its crystal structure and key physical properties, including band gap, EAH, and more.

\item \href{https://jarvis.nist.gov/}{\textcolor{black}{\textit{JARVIS-DFT} }}~\cite{choudhary2020joint}:
The Joint Automated Repository for Various Integrated Simulations (JARVIS) offers an extensive computational dataset comprising thousands of crystalline materials. Established as a component of JARVIS, JARVIS-DFT was initiated in 2017, encompassing data for approximately 40,000 materials. This dataset includes around one million calculated properties, such as crystal space groups, bandgaps, and bulk moduli. 

\item \href{http:/oqmd.org}{\textcolor{black}{\textit{OQMD} }}~\cite{saal2013materials,kirklin2015open}:
The Open Quantum Materials Database (OQMD) is a repository of thermodynamic and structural properties of inorganic materials, derived from high-throughput DFT calculations. Presently, it encompasses over 800,000 crystal structures. 

\item \href{https://github.com/txie-93/cdvae/tree/main/data/perov_5}{\textcolor{black}{\textit{Perov-5} }}~\cite{castelli2012new}:
Perov-5 is a specialized dataset designed to facilitate research on perovskite crystalline materials. It encompasses a total of 18,928 distinct perovskite materials, all sharing a common perovskite crystal structure but exhibiting compositional diversity. The dataset incorporates 56 different elements, with each unit cell containing five atoms.

\item \href{https://figshare.com/articles/dataset/Carbon24/22705192}{\textcolor{black}{\textit{Carbon-24} }}~\cite{carbon2020data}:
Carbon-24 is a specialized scientific dataset for the study of carbon materials, comprising over 10,000 distinct carbon structures that share the same elemental composition but exhibit varied structural configurations. Each entry in this dataset consists solely of carbon atoms, with unit cells containing between 6 and 24 atoms. 

\item \href{https://www.crystallography.net/}{\textcolor{black}{\textit{Crystallography Open
Database} }}~\cite{vaitkus2021validation}:
The Crystallography Open Database (COD) is a crystallography database that specializes in collecting and storing crystal structure information for inorganic compounds, small organic molecules, metal-organic compounds, and minerals. It includes specific details such as crystal structure parameters (unit cell parameters, cell volume, etc.) and space group information, allowing for the export of crystal information files in CIF format.

\item \href{https://solsa.crystallography.net/rod/index.php}{\textcolor{black}{\textit{Raman Open
Database} }}~\cite{el2019raman}:
The Raman Open Database (ROD) is an open database that specializes in collecting and storing Raman spectroscopy data. It contains a large amount of Raman spectral data for crystalline materials, including the chemical formulas of the materials and corresponding Raman spectral information such as excitation wavelengths and intensities.

\item \href{https://icsd.products.fiz-karlsruhe.de/en}{\textcolor{black}{\textit{Inorganic Crystal Structure Database} }}~\cite{bergerhoff1987crystallographic,zagorac2019recent}: 
The Inorganic Crystal Structure Database (ICSD) is the world’s largest database of fully evaluated and published crystal structure data, containing experimentally determined inorganic crystal structures as well as theoretically reported inorganic structures from the literature. The database currently includes detailed information on approximately 300,000 crystal structures, including elements, metal oxides, salts, alloys, and minerals, with details such as unit cell parameters, space groups, atomic coordinates, and more.

\item \href{https://opencatalystproject.org/}{\textcolor{black}{\textit{Open Catalyst Project} }}~\cite{zitnick2020introduction}:
The goal of Open Catalyst Project is to help artificial intelligence to simulate and discover new catalysts for renewable energy storage, aiming to help combat climate change. Currently, the Open Catalyst Project includes datasets such as Open Catalyst 2020 (OC20) and Open Catalyst 2022 (OC22), which are used to train machine learning models. These datasets collectively contain 1.3 million DFT structural relaxations (the atomic positions within the structure have been
optimized to find the configuration with the lowest energy) and results from over 260 million single-point evaluations. They cover a wide range of surfaces, and adsorbates (molecules containing nitrogen, carbon, and oxygen chemistries). Besides, these datasets are divided into training, validation, and test sets, and are used for common situations in catalysis: predicting the properties of a previously unseen adsorbate, searching for a previously unseen crystal structure or composition.

\item \href{https://doi.org/10.6084/m9.figshare.c.3938023}{\textcolor{black}{\textit{Phonon DOS Dataset} }}~\cite{petretto2018high}:
The Phonon DOS Dataset contains approximately 1,500 crystalline materials whose phonon DOS is calculated
from density functional perturbation theory. 

\item \href{http://www.carolinamatdb.org/}{\textcolor{black}{\textit{Carolina Materials Database} }}~\cite{zhao2021high}:
The Carolina Materials Database was created by groups at the University of South Carolina. The database primarily consists of ternary and quaternary materials generated by some AI methods, containing 214,436 inorganic material compounds and over 250,000 calculated properties.

\item \href{https://alexandria.icams.rub.de/}{\textcolor{black}{\textit{Alexandria Database} }}~\cite{schmidt2021crystal}: 
The Alexandria Database serves as an open-access repository, encompassing DFT-relaxed crystal structures sourced from a wide array of origins. This database includes 
a large quantity of hypothetical crystal structures generated by ML methods or other methodologies.

\item \href{https://doi.org/10.6084/m9.figshare.23713842}{\textcolor{black}{\textit{Materials Project Trajectory Dataset} }}~\cite{deng2023chgnet}:
The Materials Project Trajectory Dataset (MPtrj) comprises 1.37 million structure relaxation and static calculation tasks extracted from the MP, employing either the generalized gradient approximation (GGA) or GGA+U exchange correlation. This dataset encompasses 1,580,395 atomic configurations, corresponding energies, 7,944,833 magnetic moments, 49,295,660 force, and 14,223,555 stress.

\item \href{https://github.com/Andrew-S-Rosen/QMOF}{\textcolor{black}{\textit{Quantum MOF} }}~\cite{rosen2021machine}: Quantum MOF (QMOF) is a dataset of over 20K metal-organic frameworks (MOFs) and
coordination polymers derived from DFT. The MOFs are all DFT-optimized and are derived from a variety of parent databases, including both experimental and hypothetical MOFs databases.

\item \href{https://huggingface.co/datasets/fairchem/OMAT24}{\textcolor{black}{\textit{Open Materials 2024} }}~\cite{barroso2024open}: The Open Materials 2024 (OMat24) dataset comprises a comprehensive collection of DFT single-point calculations, structural relaxations, and molecular dynamics trajectories across a diverse array of inorganic bulk materials. In total, approximately 118 million structures have been labeled with total energy, atomic forces, and cell stress, with calculations consuming over 400 million core hours of compute resources. 

\item \href{https://github.com/pincher-chen/SODNet/tree/main/datasets/SuperCon}{\textcolor{black}{\textit{SuperCon3D} }}~\cite{chenlearning}: 
SuperCon3D utilizes crystal information, such as chemical composition, space groups, and lattice constants, to match the chemical formulas and critical temperatures ($T_c$) from the SuperCon database~\cite{stanev2018machine} with both ordered and disordered crystal structures sourced from the ICSD. 
The dataset ultimately compiles a collection of 1,578 superconductor materials, each with both $T_c$ and crystal structure data. 

\item \href{https://atomly.net}{\textcolor{black}{\textit{Atomly} }}~\cite{Atomly}: 
The Atomly database includes 320,000 inorganic crystal structures, 310,000 bandgap and density of states profiles, 12,000 dielectric constant tensors, and 16,000 mechanical tensors. 

\item \href{https://lematerial.org/}{\textcolor{black}{\textit{LeMaterial}}}~\cite{LeMaterialDatasets2025}:  
\textcolor{black}{LeMaterial is one of the largest databases. It aggregates and standardizes materials data from foundational sources such as the MP, OQMD, and Alexandria into unified datasets (e.g., LeMat-Bulk and LeMat-Traj) with consistent formats and identifiers to facilitate training and evaluation of ML models for crystalline materials.
}

\end{enumerate}

\textcolor{black}{\textbf{Benchmarks}}

\begin{enumerate}

\item \href{https://github.com/materialsproject/matbench}{\textcolor{black}{\textit{MatBench} }}~\cite{dunn2020benchmarking}: MatBench is a benchmark test suite in the field of crystalline materials, which can be used to evaluate and compare the property prediction performance of various machine learning models. The original datasets used in MatBench primarily come from multiple public databases in materials science, such as OQMD, MP, etc. Additional modifications are then enumerated to tailor the datasets for machine learning applications.

\item \href{https://github.com/yuanqidu/M2Hub}{\textcolor{black}{\textit{$M^{2}$ Hub} }}~\cite{du2023m}:
$M^{2}$ Hub is a machine learning toolkit for materials discovery research that covers the entire workflow.
In terms of datasets, $M^{2}$ Hub includes 9 public datasets, such as MP and Carbon24, covering 6 types of materials and involving 56 tasks related to eight material properties.

\item \href{https://pages.nist.gov/jarvis_leaderboard/}{\textcolor{black}{\textit{JARVIS-Leaderboard}}}~\cite{choudhary2024jarvis}:  
\textcolor{black}{JARVIS-Leaderboard is a comprehensive, reproducible benchmark platform that systematically evaluates AI, electronic structure, force-fields, quantum computation, and experimental methods for materials design across hundreds of tasks and datasets. 
}

\item \href{https://github.com/sadmanomee/OOD_Materials_Benchmark}{\textcolor{black}{\textit{Omee \textit{et al.}}}}~\cite{omee2024structure}: 
\textcolor{black}{This work presents a comprehensive benchmark for structure-based out-of-distribution (OOD) materials property prediction using GGNNs. 
The authors design five realistic OOD test generation strategies on three MatBench datasets to systematically evaluate the extrapolation capability of eight representative GNN models. 
}

\item \href{https://github.com/Bin-Cao/SimXRD}{\textcolor{black}{\textit{SimXRD-4M}}}~\cite{bin2025simxrd}:  
\textcolor{black}{SimXRD-4M introduces a open-source simulated powder X-ray diffraction (XRD) benchmark to date for crystal symmetry identification.
The dataset contains over four million high-fidelity XRD patterns generated from more than 100k crystal structures, with comprehensive multi-physical simulations that account for realistic experimental conditions.
}

\item \href{https://github.com/pincher-chen/ECDBench}{\textcolor{black}{\textit{ECD}}}~\cite{chenecd}:  
\textcolor{black}{ECD provides a large-scale benchmark for electronic charge density prediction, featuring mixed-precision PBE/HSE datasets and standardized in-distribution, transfer, experimental, and OOD evaluation protocols for crystalline materials.
}

\item \href{https://github.com/lamalab-org/chembench/}{\textcolor{black}{\textit{MaCBench}}}~\cite{alampara2025probing}:  
\textcolor{black}{MaCBench is a large-scale multimodal benchmark that evaluates vision language models on real-world chemistry and materials science tasks spanning data extraction, experiments, and data interpretation, revealing fundamental limitations in spatial reasoning and cross-modal scientific understanding.
}

\item \href{https://github.com/janosh/matbench-discovery}{\textcolor{black}{\textit{Matbench Discovery}}}~\cite{riebesell2025framework}:  
\textcolor{black}{
Matbench Discovery emphasizes prospective evaluation of thermodynamic stability prediction using unrelaxed crystal structures as input. 
The benchmark introduces discovery-relevant metrics, such as classification-based measures and discovery acceleration factor, and demonstrates that universal interatomic potentials trained on energies, forces, and stresses significantly outperform energy-only models in large-scale materials screening.
}

\end{enumerate}

\textcolor{black}{\textbf{Software Platforms}}

\begin{enumerate}
\item \href{https://pymatgen.org/}{\textcolor{black}{\textit{Python Materials Genomics} }}~\cite{ong2013python}: 
Python Materials Genomics (Pymatgen) is a robust, open-source Python library for materials analysis.
Pymatgen offer a range of modules for handling crystal structures, band structures, phase diagrams, and material properties. 
It supports a wide array of input and output formats, including VASP, ABINIT, CIF, and XYZ. 

\item \href{https://pages.nist.gov/jarvis/}{\textcolor{black}{\textit{JARVIS-Tools} }}~\cite{choudhary2020joint}: 
\textcolor{black}{JARVIS-Tools is a Python-based automation and workflow framework. It provides unified scripts and modular APIs for running, post-processing, validating, and disseminating density functional theory, classical force-field, and machine learning simulations.} 

\end{enumerate}

\bibliography{mybib}

\end{document}